\documentstyle[12pt]{article}
\newcommand{\be}{\begin{equation}}
\newcommand{\ee}{\end{equation}}
\newcommand{\ba}{\begin{eqnarray}}
\newcommand{\ea}{\end{eqnarray}}

\newcommand{\D}{{\cal{D}}}
\newcommand{\Tr}{{\rm{Tr}}}
\newcommand{\pl}{\parallel}
\date{}
\renewcommand{\theequation}{\arabic{section}.\arabic{equation}}

\newcommand{\grgl}{\:\hbox to -0.2pt{\lower2.5pt\hbox{$\sim$}\hss}
           {\raise3pt\hbox{$>$}}\:}
\newcommand{\klgl}{\:\hbox to -0.2pt{\lower2.5pt\hbox{$\sim$}\hss}
           {\raise3pt\hbox{$<$}}\:}

\begin{document}
\begin{titlepage}
\begin{flushright}
HD--THEP--94--39\\
\end{flushright}
\quad\\
\vspace{1.8cm}
\begin{center}
{\bf\LARGE Indications for Gluon Condensation}\\
\bigskip
{\bf\LARGE From Nonperturbative Flow Equations}\\
\vspace{1cm}
M. Reuter\\
\bigskip
DESY, Notkestra\ss e 85, D-22603 Hamburg \\
\vspace{.6cm}
C. Wetterich\\
\bigskip
Institut  f\"ur Theoretische Physik\\
Universit\"at Heidelberg\\
Philosophenweg 16, D-69120 Heidelberg\\
\vspace{3cm}
{\bf Abstract:}\\
\parbox[t]{\textwidth}{We employ nonperturbative flow equations for
the description of the effective action in Yang-Mills theories.
We find that the perturbative vacuum with vanishing gauge field
strength does not correspond to the minimum of the Euclidean
effective action. The true ground state is characterized by a nonvanishing
gluon condensate.}
\end{center}\end{titlepage}
\newpage

\section{Introduction}
\setcounter{equation}{0}

A perturbative calculation of the effective action for constant
colour-magnetic fields in Yang-Mills theories indicates that the
configuration
of lowest Euclidean action does not correspond to vanishing fields \cite
{sav},\cite{dit}.
This has led to many interesting speculations about the nature of the
QCD vacuum. Unfortunately, perturbation theory is clearly invalid in the
interesting region in field space. It breaks down both for vanishing
magnetic fields and for fields $B$ corresponding to the minimum of the
perturbatively calculated effective action $\Gamma[B]$. Despite many
nonperturbative indications for a nontrivial QCD vacuum from
lattice studies and phenomenology, an analytical establishment of the
phenomenon of gluon condensation is still lacking so far.

In this paper we present new analytical evidence for gluon condensation
based on the non-perturbative method of the average action \cite{Av}. The
average
action $\Gamma_k$ is the effective Euclidean action for averages of fields
which obtains by integrating out all quantum fluctuations with
(generalized)
momenta $q^2>k^2$. It can be viewed as the standard effective action
$\Gamma$ computed with an additional infrared cutoff $\sim k$ for all
fluctuations. In the limit $k\to 0$ the average action equals the
usual effective action, $\Gamma_0=\Gamma$. For $k>0$ no infrared
divergences
can appear in the computations and a lowering of $k$ allows to explore
the long-distance physics step by step. The $k$-dependence of the
average action is described by an exact nonperturbative evolution equation
\cite{Ev}.\footnote{For the relation to earlier versions of exact
renormalization group equations \cite{Rg} see ref. \cite{Mw}.}
The structure of this equation is close to a perturbative one-loop
equation
but it involves the full propagator and vertices instead of the classical
ones. For gauge theories it can be formulated in a way such that
$\Gamma_k[A]$
is a gauge-invariant functional of the gauge field $A$
\cite{reu}.\footnote{See
ref. \cite{Gau} for alternative formulations.}

The flow equation for the pure non-abelian Yang-Mills theory has been
solved
\cite{reu} with a very simple approximation - the average action
$\Gamma_k$
has been truncated with a minimal kinetic term $\sim
Z_kF_{\mu\nu}F^{\mu\nu}$. From the $k$-dependence of $Z_k$ the running of
the renormalized
gauge coupling $g(k)$ has been derived for arbitrary dimension $d$. Most
strikingly, this lowest-order estimate suggests that the wave function
renormalization $Z_k$ reaches zero for $k=k_\infty$ and turns negative
for $k<k_\infty$. Here the scale $k_\infty$ can be identified with the
confinement scale, i.e. the scale where the renormalized gauge coupling
$g^2(k)\sim Z_k^{-1}$ diverges. This was a first nonperturbative
analytical indication for the instability of the perturbative vacuum
with $F_{\mu\nu}=0$. Indeed, negative $Z_k$ implies gluon condensation -
the minimum of $\Gamma_k$ must occur for a nonvanishing gauge field
for all $k<k_\infty$ \cite{reuwe}.  It is obvious, however, that an
effective action $\sim Z_kB^2$ is insufficient to describe gluon
condensation. One expects positive $\Gamma_k$ for large $B$ and this
cannot be accomodated with a negative $Z_k$ in the ``lowest
order truncation''. In this paper we enlarge the ``space of actions''
by considering for $\Gamma_k$ an arbitrary function of constant magnetic
fields, i.e. $\Gamma_k\sim W_k(\frac{1}{2}B^2)$. The flow equation
describes how the function $W_k$ changes its shape, starting from a linear
dependence $W_k=\frac{1}{2}Z_kB^2$ for large $k$. We will see that within
our truncation $W_k$ develops its absolute minimum for $B\not=0$ once $k$
becomes smaller than a critical scale
$k_c$, and that the linear term indeed vanishes for $k_\infty
<k_c$. Despite some shortcomings of the present truncation for $k
\klgl k_\infty$ (which are discussed in the conclusions) we interpret
these findings as a strong analytical indication for gluon condensation!

In the following we shall consider the pure $SU(N)$
Yang-Mills theory. For this case the exact evolution equation of ref.
\cite{reu} reads $(t\equiv \ln k)$
\ba\label{1.1}
\frac{\partial}{\partial t}\Gamma_k[A,\bar
A]&=&\frac{1}{2}\Tr_{xcL}\left[\left(\Gamma^{(2)}_k[A,\bar
A]+R_k(\Gamma^{(2)}_k[\bar A,\bar A])\right)^{-1}\frac{\partial}{\partial
t} R_k(\Gamma^{(2)}_k[\bar A,\bar A])\right]\nonumber\\
&&-\Tr_{xc}\left[(-D^\mu[A]D_\mu[\bar A]+R_k(-D^2[\bar
A]))^{-1}\frac{\partial}{\partial t}R_k(-D^2[\bar A])\right].
\ea
Here the first trace on the r.h.s. arises from the fluctuations of the
gauge field. It involves on integration over space-time (``$x$'') as well
as a summation over color (``$c$'') and Lorentz (``$L$'') indices. The
``color'' trace is in the adjoint representation. The second trace is due
to the Faddeev-Popov ghosts and is over space-time and  color indices
only.  The precise form of the  infrared cutoff is described by the
function $R_k$. It is convenient to choose
\be\label{1.2}
R_k(u)=u[\exp({\cal Z}_k^{-1} u/k^2)-1]^{-1}.
\ee
In general the wave function renormalization
${\cal Z}_k$ could be a matrix in the space of fields which may even
depend on $\bar A$. (In ref. \cite{reu} we used ${\cal Z}_k\equiv 1$ for
the ghosts and a $k$-dependent  constant ${\cal Z}_k\equiv Z_{F,k}$ for
all modes of the gauge field.)
Furthermore, $\Gamma_k^{(2)}[A,\bar A]$ denotes the matrix of second
functional derivatives of $\Gamma_k$ with respect to $A$, with the
background field $\bar A$ kept fixed. The modes  of the gauge field are
declared ``high-frequency modes''
or ``low-frequency modes'' depending on whether  their eigenvalues with
respect to the operator  $\Gamma_k^{(2)}[\bar A,\bar
A]\equiv\Gamma_k^{(2)}[A,\bar A]\bigm\vert_{A=\bar A}$ are larger or
smaller than $k^2$, respectively. Therefore it is this operator which
appears in the argument of $R_k$\footnote{In \cite{reu} we used the
classical $S^{(2)}[\bar A,\bar A]$ rather than $\Gamma_k^{(2)}[\bar A,\bar
A]$ for this purpose. While preserving all the general properties of
$\Gamma_k$, the new flow equation is much easier to handle from a
technical point of view.  This change also entails the different
positioning of ${\cal Z}_k$ in (\ref{1.2}) relative to the one in
\cite{reu}: for the simple truncation used there one has
$\Gamma_k^{(2)}=Z_{F,k}S^{(2)}$ for part of the modes. We only will
use the definition (\ref{1.2}) if for high momenta $q^2\to\infty$
one has $u\to\infty$. We
should mention, however, that for negative eigenvalues of $\Gamma_k^{(2)}
[\bar A,\bar A]$ the vanishing of $R_k$ for $k\to0$ is guaranteed only if
the
ratio $u/k^2$ in (\ref{1.2}) remains finite. This problem concerns
mainly the approach to convexity of the effective action for $k\to0$
and is of no relevance for the present work.}
and in this sense $R_k$ acts as an effective infrared cutoff by
suppressing the low frequency modes.  The flow equation (\ref{1.1}) can be
rewritten in close analogy to a one-loop formula
\ba\label{1.3}
\frac{\partial}{\partial t} \Gamma_k[A,\bar
A]&=&\frac{1}{2}\frac{D}{Dt}\Tr_{xcL}\ln\left[\Gamma^{(2)}_k[A,\bar
A]+R_k\left(\Gamma^{(2)}_k[\bar A,\bar A]\right)\right]\nonumber\\
&&-\frac{D}{Dt}\Tr_{xc}\ln\left[-D^\mu[A]D_\mu[\bar A]+R_k(-D^2[\bar
A])\right]
\ea
The derivative $\frac{D}{Dt}$ acts only on the explicit $k$-dependence of
the function $R_k$, but not on $\Gamma^{(2)}_k [A,\bar A]$.  It is now
easy to describe the relation between the effective average action
$\Gamma_k$ and the conventional effective action. Let us first make the
approximation $\frac{D}{Dt}\to\frac{\partial}{\partial t}$ in eq.
(\ref{1.3}). This amounts to neglecting the running of $\Gamma_k$ on the
r.h.s. of the evolution equation. It is then trivial to solve it
explicitly:
\ba\label{1.4}
\Gamma_k[A,\bar A]&=&\Gamma_\Lambda[A,\bar
A]+\frac{1}{2}\Tr_{xcL}\left\{\ln\left[\Gamma_k^{(2)}[A,\bar
A]+R_k(\Gamma_k^{(2)}[\bar A,\bar A])\right]\right.\nonumber\\
&&\left.-\ln\left[\Gamma_\Lambda^{(2)}[A,\bar A]+R_\Lambda
(\Gamma_\Lambda^{(2)}[\bar A,\bar A])\right]\right\}\nonumber\\
&&-\Tr_{xc}\left\{\ln\left[-D^\mu[A]D_\mu[\bar A]+R_k(-D^2[\bar
A])\right]\right.\nonumber\\
&&\left.-\ln\left[-D^\mu[A]D_\mu[\bar A]+R_\Lambda(-D^2[\bar
A])\right]\right\}
+O\left(\frac{\partial}{\partial t}\Gamma^{(2)}_k\right)
\ea
with $\Lambda$ some appropriate high momentum scale (ultraviolet cutoff)
where we may identify $\Gamma_\Lambda$ with the classical action $S$
including a gauge fixing term. This formula has  a similar structure as a
regularized expression for the conventional one-loop effective action in
the background gauge \cite{abb,dit}. There are two important differences,
however: (i) The second variation of the classical action, $S^{(2)}$, is
replaced by $\Gamma^{(2)}_k$. This implements a kind of ``renormalization
group improvement'' and transforms (\ref{1.4}) into a sort of ``gap
equation''.
(ii)  The effective average action contains an explicit infrared cutoff
$R_k$. Because
\be\label{1.5}
\lim_{u\to\infty} R_k(u)=0,\qquad \lim_{u\to 0} R_k(u)=Z_k k^2,
\ee
a $k$-dependent mass-type term is added to the inverse propagator
$\Gamma^{(2)}_k$ for low frequency modes $(u\to 0)$, but it is absent for
high frequency modes $(u\to \infty)$. For the purpose of comparison we
will also
consider in this paper a second choice for the cutoff, namely simply
a constant
\be\label{1.6}
R_k=Z_kk^2\ee

Despite the similarity of (\ref{1.4}) with a one-loop expression, we
stress that, for $k\to 0$, the solution of the original renormalization
group equation  (\ref{1.1}) or (\ref{1.3}) equals  the {\it exact}
effective action which includes contributions  from all orders of the loop
expansion. For a detailed discussion of the approximation (\ref{1.4}) in
the case of the abelian Higgs model we refer to \cite{wet}, and to ref.
\cite{rwe} for the corresponding  nonperturbative evolution equations.

\section{Truncating the space of actions}
\setcounter{equation}{0}
In order to find nonperturbative approximative solutions of (\ref{1.1}) we
employ the following ansatz for $\Gamma_k$:
\be\label{2.1}
\Gamma_k[A,\bar A]=\int d^dx\
W_k(\frac{1}{4}F^z_{\mu\nu}(x)F^{\mu\nu}_z(x))+
\frac{1}{2\alpha_k}\int d^dx\sum_z(D_\mu[\bar A](A^\mu-\bar A^\mu))^2_z\ee
Here $W_k$ is an arbitrary function of the invariant $\frac{1}{4}F^2$ with
$F^z_{\mu\nu}$ the field strength of the gauge field $A$. The
$k$-dependence
of $W_k$ will be determined by inserting (\ref{2.1}) into the evolution
equation (\ref{1.1}). If we think of $W_k(\theta)$,
$\theta\equiv\frac{1}{4}F_{\mu\nu}^zF^{\mu\nu}_z$, as a power series
in $\theta$ our ansatz contains invariants of arbitrarily high
canonical dimension. Since all invariants which occur are of the form
$\theta^l$, only the dimensions $4l,l=1,2,...,$  actually occur. Also for
a fixed dimension $4l$ the truncation (\ref{2.1}) does not contain
a complete basis of operators. Nevertheless one may hope that (\ref{2.1})
gives a qualitatively correct picture of the physics in the regime where
the
renormalization group evolution has already drastically modified the
classical Lagrangian  $\frac{1}{4}F^2$. We remark that effective actions
which depend on $\theta$ only play also a central role in the leading-log
models \cite{adl} of QCD. The second term on the r.h.s. of (\ref{2.1}) is
a standard background gauge-fixing term \cite{abb}, \cite{reu} with a
$k$-dependent gauge-fixing parameter $\alpha_k$. In addition to the
$k$-dependence of the function $W_k(\theta)$ we should, in principle, also
compute the $k$-dependence  of $\alpha_k$. We will omit this here since
the general identities of ref. \cite{reu} imply that $\alpha$ is
independent
of $k$ in a first approximation.

We should mention at this place that a truncation is actually not defined
by the terms retained but rather by specifying which invariants in the
most general form of $\Gamma_k$ are omitted. One may parametrize a general
$\Gamma_k$ by infinitely many couplings multiplying the infinitely many
possible invariants which can be formed from the gauge fields consistent
with the symmetries. In the corrresponding infinite dimensional space
a truncation is a projection on a subspace which is defined by setting all
but
the specified couplings to zero. (In our case the subspace remains
infinite dimensional.) In practice, we will choose a particular test
configuration $A_\mu$ corresponding to a constant magnetic field. Its
detailed definition will be given below. The truncation should then be
understood in the sense that we use a basis for the invariants where all
invariants except those used in (\ref{2.1}) vanish for the test
configuration. By putting the coefficients of all invariants which
vanish for the test configuration to zero the truncation is uniquely
defined.
A computation of $W_k$ therefore amounts to a computation of the
$k$-dependent
effective action for a (particular) constant magnetic field.

Upon performing the second variation of the ansatz (\ref{2.1}),
\be\label{2.2}
\delta^2\Gamma_k[A,\bar A]=\int d^dx\delta A^\mu_y\Gamma_k^{(2)}[A,\bar A]
^{yz}_{\mu\nu}\delta A^\nu_z\ee
we arrive at the following operator $\Gamma_k^{(2)}$:
\be\label{2.3}
\Gamma_k^{(2)}[A,\bar A]^{yz}_{\mu\nu}=W_k'(\theta)({\cal D}_T[A]-{\cal D}
_L[A])^{yz}_{\mu\nu}
+W_k''(\theta){\cal S}^{yz}_{\mu\nu}[A]+\frac{1}{\alpha_k}({\cal D}_L[\bar
A])^{yz}
_{\mu\nu}\ee
with
\be\label{2.4}
\theta=\frac{1}{4}F^z_{\mu\nu}F^{\mu\nu}_z\ee
Here we used the notation ($w, y, z$ are adjoint group indices and $\bar
g$ is the (bare) gauge coupling)
\ba\label{2.5}
({\cal D}_T)^{yz}_{\mu\nu}&=&(-D^2\delta_{\mu\nu}+2i\bar g
F_{\mu\nu})^{yz}
\nonumber\\
({\cal D}_L)^{yz}_{\mu\nu}&=&-(D\otimes D)^{yz}_{\mu\nu}=-D^{yw}_\mu
D^{wz}_\nu\nonumber\\
{\cal S}^{yz}_{\mu\nu}&=&F^y_{\mu\rho}F^w_{\sigma\nu}(D^\rho
D^\sigma)^{wz}
\ea
with the covariant derivative $(D_\mu[A])^{yw}=\partial_\mu\delta^{yw}
-i\bar g A^z_\mu(T_z)^{yw}$ in the adjoint representation and
$F^{yw}_{\mu\nu}=F^z_{\mu\nu}(T_z)^{yw}$.
Moreover, $W_k'$
and $W_k''$ denote the first and the second derivative of $W_k$ with
respect
to $\theta$. In writing down eq. (\ref{2.4}) we made the additional
assumption that the field strength $F_{\mu\nu}[A]$ is covariantly
constant $F^z_{\mu\nu;\rho}=0$ or
\be\label{2.6}
[D_\rho[A],F_{\mu\nu}[A]]=0\ee
In fact, we may choose as a particularly convenient test field
a covariantly constant color-magnetic field \cite{dit} with a vector
potential of the form
\be\label{2.7}
A^z_\mu(x)=n^z{\sf A}_\mu(x)\ee
Here $n^z$ is a constant unit vector in color space $(n^zn_z=1)$, and
${\sf A}_\mu(x)$ is any ``abelian'' gauge field whose field strength
\be\label{2.8}
{\sf F}_{\mu\nu}=\partial_\mu{\sf A}_\nu-\partial_\nu{\sf A}_\mu=B\epsilon
^\perp_{\mu\nu}=const\ee
corresponds to a constant magnetic field $B$ along the 3-direction, say.
(We
define $\epsilon^\perp_{12}=-\epsilon^\perp_{21}=1$, with all other
components
vanishing.) Hence we have $\theta=\frac{1}{4}F^z_{\mu\nu}F^{\mu\nu}_z=
\frac{1}{2}B^2$ such that $\theta$ and $W_k'$ etc. commute with all
operators. It is easy to see that (\ref{2.7}) and (\ref{2.8}) obey the
condition (\ref{2.5}). The choice (\ref{2.7}) has the advantage that it
allows for an explicit diagonalization of the operator $\Gamma_k^{(2)}$.
We
finally note that $W_k$ can be extracted from $\Gamma_k[A,A]$ which is a
gauge-invariant functional of $A$ obtained by putting $\bar A=A$. It is
therefore sufficient to know $\Gamma_k^{(2)}[A,A]$.

Before returning to the evolution equation, we list a few special
properties
of the covariantly constant fields, which will prove helpful later on.
{}From (\ref{2.5})
it follows that $A_\mu$ satisfies the classical Yang-Mills equations
$D^\mu
F_{\mu\nu}=0$. This in turn is sufficient to prove that the operators
${\cal D}_T$ and ${\cal D}_L$ commute. As a consequence, one may define
generalized projection operators \cite{reu}
\ba\label{2.9}
P_L&=&{\cal D}^{-1}_T{\cal D}_L\nonumber\\
P_T&=&1-P_L\ea
which satisfy $P_{T,L}^2=P_{T,L},P_T+P_L=1$ and $P_TP_L=0=P_LP_T$. For
$A_\mu=0$ they reduce to the standard projectors on transverse and
longitudinal modes:
\ba\label{2.10}
(P_T^{(0)})_{\mu\nu}&=&\delta_{\mu\nu}-\partial_\mu\partial_\nu/\partial^2
\nonumber\\
(P_L^{(0)})_{\mu\nu}&=&\partial_\mu\partial_\nu/\partial^2\ea
Furthermore, if $A_\mu$ is of the form (\ref{2.7}), it is natural
to define another pair of orthogonal projectors,
\be\label{2.11}
P_\perp^{yz}=\delta^{yz}-n^yn^z,\quad P_{\parallel}^{yz}=n^yn^z,\ee
which project on the spaces perpendicular and parallel to $n^z$,
respectively. For the vector potential (\ref{2.7}) the matrix
$A^w_\mu T_w$ reads in the adjoint representation
\be\label{2.12}
A^{yz}_\mu(x)\equiv(A^w_\mu T_w)^{yz}=if^{ywz}n_w{\sf A}_\mu(x)\ee
The antisymmetry of the structure constants $f^{ywz}$ implies that
$P_{\parallel}$ and $P_\perp$ commute with $D_\mu,D^2,{\cal D}_T, {\cal D}
_L$ and $F_{\mu\nu}$, and that
\be\label{2.13}
P_{\parallel}A_\mu=0,\quad P_\parallel D_\mu=P_\parallel
\partial_\mu,\quad
P_\parallel({\cal D}_T)_{\mu\nu}=-\partial^2\delta_{\mu\nu}P_\parallel\ee
The operator ${\cal S}$ from (\ref{2.5}) factorizes according to
\ba\label{2.14}
{\cal S}^{yz}_{\mu\nu}&=&P^{yz}_{||}s_{\mu\nu},\nonumber\\
s_{\mu\nu}&=&{\sf F}_{\mu\rho}{\sf
F}_{\sigma\nu}\partial^\rho\partial^\sigma
\ea
Hence ${\cal S}$ commutes with $D^2, {\cal D}_L$ and ${\cal D}_T$ because
it annihilates the gauge-field
contained in these operators:
\be\label{2.15}
{\cal S}D^2={\cal S}\partial^2,\quad {\cal S}{\cal D}_T=-{\cal
S}\partial^2,
\quad {\cal S} D\otimes D={\cal S}\partial\otimes\partial\ee

In physical terms this means that those components of the gauge
fluctuations $\delta A^z_\mu\equiv a^z_\mu$ which are parallel
to $n^z$ decouple from $A_\mu$ to some extent. In fact, in terms of the
projections $a_\mu^{\perp,||}=P_{\perp,||}a_\mu$ the quadratic action
(\ref{2.2})
with (\ref{2.6}) reads
\ba\label{2.16}
&&\delta^2\Gamma_k[A,A]=\int d^dx\left\lbrace a^{||\mu}_z[-\partial^2
W_k'\delta_{\mu\nu}+(W_k'-\frac{1}{\alpha_k})\partial_\mu\partial_\nu
\right.
+W_k''s_{\mu\nu}]a^{||\nu z}\nonumber\\
&&\left.+a^{\perp \mu}_y[W_k'{\cal D}_T+(\frac{1}{\alpha_k}-W_k'){\cal
D}_L]^{yz}_{\mu\nu}a^{\perp\nu}_z\right\rbrace\ea
We observe that the $a^{||}$-modes couple to the external field only via
the
derivatives of $W_k\equiv W_k(\frac{1}{2}B^2)$. In a conventional
one-loop calculation one uses the classical Yang-Mills Lagrangian
$\frac{1}{4}F^2_{\mu\nu}$ rather than $W_k(\frac{1}{4} F^2_{\mu\nu})$.
In that case the quadratic action for the small fluctuations is given by
(\ref{2.16}) with $W_k'=1$ and $W_k''=0$. Hence the one-loop determinant
resulting from the integration over $a^{||}$ is field-independent and may
be ignored. In the present case, the $a^{||}$-modes are important for the
``renormalization group improvement'', however. The quadratic form
(\ref{2.16}) can be diagonalized even further by introducing the
longitudinal and transversal projections
\ba\label{2.17}
&&a^{||,L}=P_La^{||}=P_L^{(0)}a^{||},\quad
a^{||,T}=P_Ta^{||}=P_T^{(0)}a^{||}
\nonumber\\
&&a^{\perp,L}=P_L a^\perp,\quad a^{\perp,T}=P_Ta^\perp\ea
By virtue of $P_L^{(0)}s=sP_L^{(0)}=0,P_T^{(0)}s=sP_T^{(0)}=s$
and $[P_{L(T)},\theta]=0$ one obtains
\ba\label{2.18}
\delta^2\Gamma_k[A,A]&=&\int d^dx\left\lbrace a^{||,T,\mu}_z[-\partial^2
W_k'\delta_{\mu\nu}
+W_k''s_{\mu\nu}]a^{||,T,\nu,z}\right.\nonumber\\
&&+\frac{1}{\alpha_k}a^{||,L,\mu}_z[-\partial^2]a^{||,L,z}_\mu\nonumber\\
&&+a^{\perp,T,\mu}_y[W_k'{\cal
D}_T]^{yz}_{\mu\nu}a^{\perp,T,\nu}_z\nonumber\\
&&\left.+\frac{1}{\alpha_k}a^{\perp,L,\mu}_y[{\cal
D}_T]^{yz}_{\mu\nu}a^{\perp,L,\nu}_z\right\rbrace\ea
This block-diagonal form of $\Gamma_k^{(2)}$ will facilitate the evolution
of the traces occurring in the evaluation equation. For example,
$a^{||,L}$
gives no $A$-dependent contribution and, except for an irrelevant
constant,
the only dependence of $\Gamma_k$ on $\alpha_k$ arises from $a^{\perp,L}$.
Writing
\be\label{2.19}
\Gamma^{(2)}[A,A]=\Gamma_1^{(2)}+\Gamma_2^{(2)}+\Gamma_3^{(2)}+
\Gamma_4^{(2)}\ee
where, in an obvious notation
\ba\label{2.20}
&&\Gamma_1^{(2)}=P_{||}P_T\Gamma^{(2)}_{||,T}P_{||}P_T,\quad
\Gamma_2^{(2)}=P_{||}P_L\Gamma^{(2)}_{||,L}P_{||}P_L\nonumber\\
&&\Gamma_3^{(2)}=P_{\perp}P_T\Gamma^{(2)}_{\perp,T}P_{\perp}P_T,
\quad\Gamma_4^{(2)}=P_{\perp}P_L\Gamma^{(2)}_{\perp,L}P_{\perp}P_L\ea
with
$[P_{||,\perp},P_{L,T}]=0$ and
\ba\label{2.21}
&&\Gamma_A^{(2)}\Gamma_B^{(2)}=0\quad{\rm for}\quad A\not=B\nonumber\\
&&[\Gamma_A^{(2)},\Gamma_B^{(2)}]=0\ea
one obtains ([$\Gamma^{(2)}_{||,T},P_{||}P_T]=0$ etc.)
\ba\label{2.22}
R_k(\Gamma^{(2)})&=&P_{||}P_TR_k(\Gamma_{||,T}^{(2)})P_{||}P_T
+P_{||}P_LR_k(\Gamma_{||,L}^{(2)})P_{||}P_L\nonumber\\
&&+P_{\perp}P_TR_k(\Gamma_{\perp,T}^{(2)})P_{\perp}P_T
+P_{\perp}P_LR_k(\Gamma_{\perp,L}^{(2)})P_{\perp}P_L\ea

\section{Evolution equation for $W_k$}
\setcounter{equation}{0}
In this section we derive an evolution equation which governs the
scale-dependence of the function $W_k$. We will choose the matrix
${\cal Z}_k$ in the
definition of $R_k$ (\ref{1.2}) as ${\cal Z}_k=1$ for the ghosts and
\be\label{3.1}
{\cal Z}_k=Z_kP_T[\bar A]+\tilde Z_kP_L[\bar A]\ee
for the gauge boson degrees of freedom. Here $Z_k,\tilde Z_k$ are
$k$-dependent constants and we observe that the choice (\ref{3.1}) is
compatible with (\ref{2.22}). If we insert the truncation (\ref{2.1}) into
(\ref{1.1}) with $\bar A=A$, we obtain $(\theta=\frac{1}{2}B^2)$
\ba\label{3.2}
\Omega\frac{\partial}{\partial t}W_k(\theta)&=&\frac{1}{2}{\rm
Tr}_{xcL}[H(
\Gamma_k^{(2)}[A,A])]\nonumber\\
&&-{\rm Tr}_{xc}[H_G(-D^2[A])]\nonumber\\
&&+\frac{1}{2}{\rm Tr}_{xcL}[P_\perp P_L(\tilde H(\Gamma_{\perp
L}^{(2)})-H(\Gamma_{\perp
L}^{(2)}))]\ea
where $\Gamma_k^{(2)}$ is given by (\ref{2.6}) and $\Omega\equiv \int
d^dx$.
In (\ref{3.2}) we introduced the convenient abbreviation
(for ${\cal Z}_k=Z_k$)
\ba\label{3.3}
H(u)&\equiv&(u+R_k(u))^{-1}\frac{\partial}{\partial t}R_k(u)\nonumber\\
&=&\left\lbrace
\begin{array}{lll}
(2+\frac{d}{dt}\ln
Z_k)\frac{u}{Z_kk^2}\left[\exp\left(\frac{u}{Z_kk^2}\right)-1\right]^{-1}&
{\rm for}& (1.2)\\
(2+\frac{d}{dt}\ln Z_k)Z_kk^2[u+Z_kk^2]^{-1}&{\rm for}&(1.6)\end{array}
\right.
\ea
In the second, ghost-type, trace the function $H_G(u)$ is defined
similarly,
but with a different factor ${\cal Z}_k=1$. The last trace accounts for a
possible difference between $\tilde Z_k$ and $Z_k$ in (\ref{3.1}).
Here $\tilde H$ is obtained from $H$ by replacing $Z_k$ by $\tilde Z_k$
and
for $\Gamma_{\perp L}^{(2)}$ we may take equivalently
${\cal D}_T/\alpha_k$ or ${\cal D}_L/\alpha_k$.

In appendix A we display various trace identities for an evaluation of
(\ref{3.2}). They are exploited in appendix B to express the evolution
equation for $W_k$ as traces over simple operators. We note that the
ultraviolet finiteness of the momentum integrals implied by the trace
is guaranteed for the choice (\ref{1.2}) only if $H(u)$ decays
exponentially for large $q^2$. This requires $W_k'>0,W_k'+B^2W_k''>0$.
Restricting the discussion to this case, one finds
\ba\label{3.4}
\frac{\partial}{\partial t}W_k(\frac{1}{2}B^2)&=&\frac{1}{2}\Omega^{-1}
\Tr_{xcL}[H(W_k'{\cal D}_T)]\nonumber\\
&&+\frac{1}{2}\Omega^{-1}
\Tr_{xc}[\tilde H(-\alpha_k^{-1}D^2)-H(-W_k' D^2)]
-\Omega^{-1}\Tr_{xc}[H_G(-D^2)]\nonumber\\
&&+v_d\left(\frac{1}{W_k'+B^2W_k''}-\frac{1}{W_k'}\right)
\left(\frac{1}{W_k'}
\right)^{\frac{d}{2}-1}\int^\infty_0dx x^{\frac{d}{2}-1}H(x)\nonumber\\
&&-v_d\int^\infty_0dx x^{\frac{d}{2}-1}(\tilde
H\left(\frac{x}{\alpha_k}\right)
-H\left(\frac{x}{\alpha_k}\right))\ea
with
\be\label{3.5}
v_d^{-1}=2^{d+1}\pi^{\frac{d}{2}}\Gamma\left(\frac{d}{2}\right)\ee
The eigenvalues of the operator ${\cal D}_T$ are known explicitly
\cite{dit}.
They are parametrized by a $(d-2)$-dimensional momentum $q^\mu$ which
``lives'' in the space orthogonal to the 1-2 plane, and a discrete quantum
number $n=0,1,2,...$ which labels the Landau levels. The spectral sum
for the function $\hat H(x)\equiv H(W_k'x)$ reads
\ba\label{3.6}
\Omega^{-1}\Tr_{xcL}[\hat H({\cal D}_T)]&=&\sum^{N^2-1}_{l=1}\frac{\bar
g|\nu_l|B}{2\pi}\sum^\infty_{n=0}\int
\frac{d^{d-2}q}{(2\pi)^{d-2}}\nonumber\\
&&\cdot\left\lbrace (d-2)\hat H(q^2+(2n+1)\bar
g|\nu_l|B)\right.\nonumber\\
&&+\hat H (q^2+(2n+3)\bar g|\nu_l|B)\nonumber\\
&&\left.+\hat H (q^2+(2n-1)\bar g|\nu_l|B)\right\rbrace\ea
(The momentum integration is absent for $d=2$.)
Here $\nu_l,l=1,..., N^2-1$ are the eigenvalues of the matrix $n^zT_z$ in
the adjoint representation. We note that for $n=0$ and $q^2$ sufficiently
small the eigenvalue $q^2-\bar g|\nu_l|B$ in the third term on the r.h.s.
of
(\ref{3.6}) can become negative. This is the origin of the instability
\cite{nol} of the Savvidy vacuum \cite{sav}, which causes severe problems
if one tries to compute the standard one-loop effective action in
the background of a covariantly constant magnetic field \cite{ksw}.
In our approach
this problem is cured by the presence of an IR regulator. For $d>2$
eq. (\ref{3.6}) can be rewritten as
\ba\label{3.7}
&&\Omega^{-1}\Tr_{xcL}[\hat H({\cal D}_T)]=\frac{v_{d-2}}{\pi}\sum^{N^2-1}
_{l=1}\bar g|\nu_l|B\int^\infty_0dx x^{\frac{d}{2}-2}\nonumber\\
&&\cdot \left\lbrace d\sum^\infty_{n=0}\hat H(x+(2n+1)\bar g|\nu_l|B)+\hat
H(x-\bar g|\nu_l|B)-\hat H(x+\bar g|\nu_l|B)\right\rbrace\ea
but it cannot be simplified any further in closed form. The other traces
in (\ref{3.4})
are given by
\be\label{3.8}
\Omega^{-1}\Tr_{xc}[\hat H(- D^2)]=\frac{v_{d-2}}{\pi}\sum^{N^2-1}
_{l=1}\bar g|\nu_l|B\int^\infty_0dx x^{\frac{d}{2}-2}
\sum^\infty_{n=0}\hat H(x+(2n+1)\bar g|\nu_l|B)\ee
and similar for $\tilde H$ and $H_G$.

In this paper we use two different methods in order to (approximately)
compute
the spectral sums (\ref{3.7}) and (\ref{3.8}). In appendix E we shall
represent
them as Schwinger proper-time integrals \cite{direu}. This method leads to
compact
integral representations which are valid for all values of $B$, but
it has the disadvantage that it works only for a slightly simplified
form of the cutoff function $R_k(x)$ defined by (\ref{1.6}) which leads,
as
we shall see, to ultraviolet problems.
The second method consists
of expanding the r.h.s. of (\ref{3.7}) in powers of $B$. It is applicable
if $\bar g B\ll k^2$, and it works for any function $R_k(x)$. The
condition $\bar g B\ll k^2$ guarantees that we may express the sum
over $n$ by an Euler-McLaurin series, and that the first terms
converge rapidly. In this
manner (\ref{3.7}) turns into
\ba\label{3.9}
&&\Omega^{-1}\Tr_{xcL}[H(W_k'(\frac{1}{2}B^2){\cal D}_T)]\nonumber\\
&&=(N^2-1)\frac{dv_{d-2}}{2\pi}[W_k'(\frac{1}{2}B^2)]^{-\frac{d}{2}}
\int^\infty_0 dx\int^\infty_0dy x^{\frac{d}{2}-2}H(x+y)\\
&&+\frac{v_{d-2}}{\pi}\sum^\infty_{m=1}C^d_m\left(\sum^{N^2-1}_{l=1}
\nu^{2m}_l\right)(\bar g
B)^{2m}[W_k'(\frac{1}{2}B^2)]^{2m-\frac{d}{2}}\int^\infty_0dx\
x^{\frac{d}{2}
-2}H^{(2m-1)}(x)\nonumber\ea
with
\be\label{3.10}
C^d_m=\frac{d}{(2m)!}(2^{2m-1}-1)B_{2m}-\frac{2}{(2m-1)!}\ee
Here $B_{2m}$ are the Bernoulli numbers. In a second step one has to
expand
the $B^2$-dependence of $W_k'$. Note that only even powers of $B$ occur in
this
expansion. The group-theoretical factors $\sum^{N^2-1}_{l=1}\nu^{2m}_l$
are discussed in appendix C. In particular for
$SU(2)$ these factors equal 2 for all values of $m$.

If we use the Euler-McLaurin series (3.9) and a similar expansion for the
trace (\ref{3.8}) in eq. (\ref{3.4}), we find for $SU(N)$:
\ba\label{3.11}
&&\frac{\partial}{\partial t}W_k(\theta)=\frac{v_{d-2}}{2\pi}(2-\eta)k^d
\left\lbrace\frac{d-1}{2}(N^2-1)r^d_2\left(\frac{W_k'}
{Z_k}\right)^{-\frac{d}{2}}\right.\nonumber\\
&&\left.-\sum^\infty_{m=1}\tau_m(C^d_m-E_m)r^{d,m}_0\left(\frac
{2\bar
g^2\theta}{k^4}\right)^m\left(\frac{W_k'}{Z_k}\right)^{2m-\frac{d}{2}}
\right\rbrace\nonumber\\
&&+\frac{v_{d-2}}{2\pi}(2-\tilde \eta)(\tilde
Z_k\alpha_k)^{\frac{d}{2}}k^d\left\lbrace\frac{1}{2}(N^2-1)r^d_2\right.
\left.-\sum^\infty_{m=1}\tau_m E_mr^{d,m}_0\left(\frac
{2\bar g^2\theta}{k^4}\right)^m(\tilde Z_k\alpha_k)^{-2m}
\right\rbrace\nonumber\\
&&-\frac{v_{d-2}}{\pi}k^d\left\lbrace(N^2-1)r^d_2-2\sum^\infty_{m=1}
\tau_m E_mr^{d,m}_0\left(\frac
{2\bar g^2\theta}{k^4}\right)^m
\right\rbrace\nonumber\\
&&+v_d(2-\eta)r^d_1k^d\left(\frac{Z_k}{W_k'+2\theta
W_k''}-\frac{Z_k}{W_k'}\right)\left(\frac{W_k'}{Z_k}
\right)^{1-\frac{d}{2}}
+\ {\rm const.}\ea
with $\tau_m$ defined in appendix C and the
constants $E_m$ given by
\be\label{3.12}
E_m=\frac{1}{(2m)!}(2^{2m-1}-1)B_{2m}.\ee
The dimensionless integrals
\ba\label{3.13}
r^{d,m}_0&=&-\frac{1}{2-\eta}(Z_kk^2)^{2m-\frac{d}{2}}
\int^\infty_0dxx^{\frac{d}{2}-2}H^{(2m-1)}(x)\nonumber\\
&=&-\int^\infty_0dx\ x^{\frac{d}{2}-2}\left(\frac{d}{dx}\right)^{2m-1}
\frac{x}{e^x-1}\nonumber\\
r^{d}_1&=&\frac{1}{2-\eta}(Z_kk^2)^{-\frac{d}{2}}
\int^\infty_0dxx^{\frac{d}{2}-1}H(x)=\int^\infty_0dx\
\frac{x^{\frac{d}{2}}}{e^x-1}\nonumber\\
r^d_2&=&\frac{1}{2-\eta}(Z_kk^2)^{-\frac{d}{2}}\int^\infty_0dx
\int^\infty_0dyx^{\frac{d}{2}-2}H(x+y)\nonumber\\
&=&
\int^\infty_0dx\int^\infty_0dy\frac{x^{\frac{d}{2}-2}(x+y)}{\exp(x+y)-1}
\ea
occur as a consequence of (\ref{3.3}) and the second equality uses
(\ref{1.2}).
For this choice we note for  later use  that in
4 dimensions
\be\label{3.14}
r^{4,m}_0=B_{2m-2},\ r^4_1=r^4_2=2\zeta(3)\ee
where $\zeta$ denotes the Riemann zeta function. In contrast,
the quantities $r^4_1$ and $r^4_2$ are not well defined for the choice
(\ref{1.6}). The ultraviolet divergence indicates an incomplete ``thinning
out'' of the high momentum degrees of freedom for a simple mass like
infrared cutoff. Even though eq. (\ref{3.11})
was derived by choosing a specific background field, this
evolution equation  does not depend on the background we used for the
calculation. By employing derivative expansion techniques \cite{zuk},
\cite{mgs} it should also be possible to derive (\ref{3.11}) without ever
specifying a  background. In the case at hand the method presented here
is by far simpler, however.

In eq. (\ref{3.11}) we have introduced the anomalous dimensions
\be\label{3.15}
\eta=-\frac{d}{dt}\ln Z_k,\quad\tilde\eta=-\frac{d}{dt}\ln \tilde Z_k\ee
A convenient choice for the wave function renormalization constants
used in the infrared cutoff $R_k$ is ($k_{np}$ denotes the transition
to the nonperturbative regime)
\ba\label{3.16}
Z_k&=&\left\lbrace\begin{array}{ccc}
W_k'(0)&\ {\rm for}&\ k>k_{np}\\
W_{k_{np}}'(0)&\ {\rm for}&\ k<k_{np}\end{array}\right.\nonumber\\
\tilde Z_k&=&\frac{1}{\alpha_k}\ea
Here we account for the possibility that $W_k'(0)$ may turn negative for
$k$ smaller than a ``confinement scale'' $k_\infty$, whereas $Z_k$ must
always be strictly positive.  More precisely, if $|d\ln W_k'(0)/dt|$ grows
larger than one for small scales $k$, we
choose $k_{np}$ to be the scale where
$\eta(k_{np})=1$. For $k<k_{np}$ all couplings run fast anyhow,
and an improvement of the scaling properties of $R_k$ by the introduction
of a $k$-dependent wave function renormalization seems not necessary.
 The choice $\tilde Z_k=\alpha_k^{-1}$ guarantees
that the infrared cutoff acts on the longitudinal modes in the same way
as on the transversal modes. It implies that the flow equation for
$\frac{\partial}{\partial t} W_k$ becomes independent of $\alpha_k$ in our
truncation (except for an irrelevant constant)! In this paper we can
neglect the running of $\alpha_k$. This can be infered from a first-order
approximation to the solution of the general identities which govern
the dependence of $\Gamma_k[A,\bar A]$ on the background field
$\bar A[1]$. On therefore has $\tilde\eta=0$.

It is also convenient to express the flow equation in terms of
renormalized
dimensionless quantities
\ba\label{3.17}
g^2&=&k^{d-4}Z^{-1}_k\bar g^2\nonumber\\
\vartheta&=&g^2k^{-d}Z_k\theta\nonumber\\
w_k(\vartheta)&=&g^2k^{-d}W_k(\theta)\ea
Switching to a notation where dots denote derivatives with respect
to $\vartheta$ instead of $\theta$ and $\partial/\partial t$ is now
taken at fixed $\vartheta$ one obtains
\ba\label{3.18}
&&\frac{\partial}{\partial t}w_k(\vartheta)=-(4-\eta)w_k(\vartheta)+
4\vartheta\dot w_k(\vartheta)\nonumber\\
&&+(2-\eta)v_dg^2(\dot w_k(\vartheta))^{-\frac{d}{2}}\left\lbrace
\frac{(d-1)(d-2)}{2}(N^2-1)r^d_2\right.\nonumber\\
&&\left.-\frac{2r^d_1\vartheta\ddot w_k(\vartheta)}{\dot w_k(\vartheta)+
2\vartheta\ddot w_k(\vartheta)}-(d-2)\sum^\infty_{m=1}\tau_m
(C^d_m-E_m)r^{d,m}_0(2\vartheta\dot
w_k^2(\vartheta))^m\right\rbrace\nonumber\\
&&+2(d-2)v_dg^2\sum^\infty_{m=1}\tau_mE_mr^{d,m}_0(2\vartheta)^m+{\rm
const}
\ea
This is a nonlinear partial differential equation for the function
$w_k(\vartheta)$ depending on $k$ and $\vartheta$. One needs in
addition the running of the renormalized gauge coupling $g$ and $\eta$,
which are related by
\be\label{3.19}
\beta_{g^2}=\frac{\partial g^2}{\partial t}=(d-4+\eta)g^2\ee
For $k>k_{np}$ we have by definition $\dot w_k(0)=1$ and $\eta$ can be
determined by
\ba\label{3.20}
&&\frac{\partial}{\partial t}\dot
w_k(0)=0=\eta-2(d-2)v_dr^{d,1}_0\tau_1g^2\left((2-\eta)C^d_1-(4-\eta)E_1
\right)\nonumber\\
&&-(2-\eta)v_dg^2\left(2r^d_1+\frac{d(d-1)(d-2)}{4}(N^2-1)r^d_2\right)
\ddot w_k(0)\ea
With $\tau_1=N,C^d_1=\frac{d}{12}-2,E_1=\frac{1}{12}$ this yields
\ba\label{3.21}
\eta=&-&\left(\frac{N}{3}v_d(d-2)(26-d)r^{d,1}_0g^2-2h_dg^2w_2\right)
\nonumber\\
&&\left(1-\frac{N}{6}v_d(d-2)(25-d)r^{d,1}_0g^2+h_dg^2w_2\right)^{-1}
\ea
with
\be\label{3.22}
h_d=v_d\left(2r^d_1+\frac{d(d-1)(d-2)}{4}(N^2-1)r^d_2\right)\ee
\be\label{3.23}
w_2=\ddot w_k(0)\ee
For general $d$ the constants $r^{d,1}_0,r^d_1$ and $r^d_2$ depend
on the precise choice of the infrared cutoff except
for $d=4$ where $r^{4,1}_0=1$ is cutoff independent. The running of the
renormalized gauge coupling $g$ is now determined (cf. (\ref{3.19})). It
depends on the additional coupling $w_2$ (\ref{3.23}) which will be
discussed
in more detail in the next section\footnote{For $w_2=0$ we recover the
result
of ref. \cite{reu} except for the factor $(25-d)$ in the denominator
in (\ref{3.21}) which was $(24-d)$ previously. This difference is due to
a slightly different choice of the $Z$-factors in the infrared cutoff.}.
Specifying the initial value $g^2(\Lambda)$ and the function
$w_\Lambda(\vartheta)$ at some high momentum scale $\Lambda$ the form of
$w_k(\vartheta)$ and $g^2(k)$ are completely determined by the flow
equation (\ref{3.18}). Solving for $k\to 0$ the function
$w_0(\vartheta)$ specifies the effective action in our truncation. If
necessary, one has to replace for $k<k_{np}$ eq. (\ref{3.21}) by
$\eta=0,g^2(k<k_{np})=g^2(k_{np})$.

Before closing this section, we briefly comment on the range of
convergence
of the Euler-McLaurin series in our case. For large $m$ one has
\be\label{3.24}
\lim_{m\to\infty}C^d_m=d\lim_{m\to\infty}E_m\sim\pi^{-2m}\ee
For $N=2$ and $d=4$ we find (cf. (\ref{3.14})) that the coefficients
of $(2\vartheta)^m$ in eq. (\ref{3.18}) diverge $\sim\pi^{-2m}B_{2m-2}
\sim\pi^{-4m}2^{-2m}(2m-4)!$. For small nonvanishing $\vartheta$ the
first terms of the series have a very rapid apparent convergence,
but the series finally diverges due to the factorial growth $\sim
(2m-4)!$. We can therefore safely use this series only for the derivatives
$w^{(n)}(\vartheta=0)$ with finite $n$ where convergence problems are
absent since only a finite number of terms in the sum contributes.
The situation is probably similar for $N>2$ and/or $d\not=4$ as well
as for many other choices of the infrared cutoff. In contrast, the
original sums over $n$ in eqs. (\ref{3.7}), (\ref{3.8}) always converge
since for $B>0$ the contributions from sufficiently high values of $n$
are exponentially suppressed. An explicit evaluation of these sums
is possible for the simplified masslike IR-cutoff (\ref{1.6}). This
is described in appendix E where we will also see the reason for the
ultraviolet divergence of $r_1^4$ and $r^4_2$ for this particular cutoff.

\section{The $F^4$ approximation}
\setcounter{equation}{0}
In order to get a first idea of the physical contents of the evolution
equation (\ref{3.11}) we further simplify the problem in this section:
We include in $W_k(\theta)$ only terms which are at most quadratic
in $\theta$:
\ba\label{4.1}
W_k(\theta)&=&W_0(k)+W_1(k)\theta+\frac{1}{2}W_2(k)\theta^2,\nonumber\\
&&W_j(k)\equiv \left(\frac{d}{d\theta}\right)^jW_k(0)\ea
Thus the truncation for $\Gamma_k$ is parametrized by three couplings:
\be\label{4.2}
\Gamma_k[A,A]=\int d^dx\left\lbrace
W_0(k)+\frac{1}{4}W_1(k)F^z_{\mu\nu}F^{\mu\nu}_z
+\frac{1}{32}W_2(k)\left(F^z_{\mu\nu}F^{\mu\nu}_z\right)^2\right\rbrace
\ee
Apart from the constant $W_0$, we keep here only
the ``marginal'' coupling $W_1$ and the lowest ``irrelevant'' coupling
consistent with the truncation, $W_2$. We concentrate on $d=4$ and employ
the infrared cutoff (1.2) with (\ref{3.13}). The short distance or
``classical'' theory is specified for $k=\Lambda$ by $(Z_\Lambda=1)$
\be\label{4.3}
W_1(\Lambda)=1,\quad W_0(\Lambda)=W_2(\Lambda)=0\ee
In terms of the couplings $w_0=w(\vartheta=0), g^2$ and $w_2$
(\ref{3.17}),
(\ref{3.23}), one has for $k>k_{np}$
\ba\label{4.4}
W_0(k)&=&k^4w_0(k)/g^2(k)\nonumber\\
W_1(k)&=&\bar g^2/g^2(k)\nonumber\\
W_2(k)&=&\bar g^4w_2(k)/g^2(k)k^4\ea
or
\be\label{4.5}
g^2(\Lambda)=\bar g^2,\quad w_0(\Lambda)=w_2(\Lambda)=0\ee
with $g^2(k)$ determined by eqs. (\ref{3.19}), (\ref{3.21}). For
$k<k_{np}$
we use $g^2(k)=g^2(k_{np})$ and
\ba\label{4.6}
W_1(k)&=&\frac{\bar g^2}{g^2(k_{np})}w_1(k)\nonumber\\
w_1(k)&=&\dot w_k(0)\ea
The flow equations for $w_j(k)=w_k^{(j)}(\vartheta=0)$ follow from
appropriate partial differentiation of eq. (3.18) with respect to
$\vartheta$,
as, for example
\ba\label{4.7}
&&\frac{\partial}{\partial t}\dot w=\eta\dot w+4\vartheta\ddot
w\nonumber\\
&&-(2-\eta)v_dg^2\dot w^{-\frac{d}{2}}\left\lbrace (d-2)\sum^\infty_{m=1}
\tau_m(C^d_m-E_m)\right. r^{d,m}_0(2\vartheta\dot w^2)^{m-1}(2m\dot
w^2+(4m-d)\vartheta\dot w\ddot w)\nonumber\\
&&+r^d_1\left(\frac{2\ddot w+2\vartheta w^{(3)}}{\dot w+2\vartheta\ddot
w}-
\frac{2\vartheta\ddot w(3\ddot w+2\vartheta w^{(3)})}{(\dot
w+2\vartheta\ddot w
)^2}-d\frac{\vartheta(\ddot w)^2}{\dot w(\dot w+2\vartheta\ddot w)}\right)
\nonumber\\
&&\left.+\frac{1}{4}d(d-1)(d-2)(N^2-1)r^d_2\frac{\ddot w}{\dot w}
\right\rbrace\nonumber\\
&&+4(d-2)v_dg^2\sum^\infty_{m=1}m\tau_mE_mr^{d,m}_0(2\vartheta)^{m-1}
\ea
We observe that $w_0$ does not appear on the r.h.s. of the evolution
equations for $w_j, j\geq1$, and we omit in the following this irrelevant
constant. For $k>k_{np}$ the evolution equation for the running gauge
coupling reads $(d=4)$
\be\label{4.8}
\frac{\partial g^2}{\partial t}=-\frac{g^4}{24\pi^2}\left(11N-
3H_4w_2\right)\left[1-\frac{g^2}{32\pi^2}
\left(7N-2H_4w_2\right)\right]^{-1}\ee
whereas for $k<k_{np}$ one uses the flow equation for $w_1$
\be\label{4.9}
\frac{\partial}{\partial
t}w_1=\frac{g^2(k_{np})}{8\pi^2}\left(\frac{11N}{3}
-H_4\frac{w_2}{w_1^3}\right)\ee
where $H_4=16\pi^2h_4=r^4_1+3(N^2-1)r^4_2=2(3N^2-2)\zeta(3)$
for the choice (\ref{1.2}). It is obvious that for $w_2<0$ the gauge
coupling $g^2(k)$ always increases (\ref{4.8}) until at $k=k_{np}$ the
anomalous dimension $\eta$ reaches one. If $w_2$ remains negative for
$k<k_{np}$, the coupling $w_1$ decreases until it reaches zero at the
confinement scale $k_\infty>0$. (If we define $g^2(k)=g^2(k_{np})/w_1$,
this
coupling diverges at the confinement scale.)
The issue is different for $w_2>0$: The coupling $w_1$ would not reach
zero
since for small enough $w_1$ the second term in eq. (\ref{4.9}) would
cancel
the first term. One therefore needs an estimate of $w_2(k)$.

Let us first consider $k>k_{np}$ where the evolution equation for $w_2$
reads
(with $E_2=-\frac{7}{720},\ C^4_2=-\frac{67}{180}$)
\ba\label{4.10}
&&\frac{\partial}{\partial
t}w_2=(4+\eta)w_2+\frac{g^2}{8\pi^2}\left\lbrace
\tau_2r_0^{4,2}\left(\frac{127}{45}-\frac{29}{20}\eta\right)\right.\\
&&\left.+(2-\eta)\left(5r^4_1+\frac{9}{2}(N^2-1)r^4_2\right)
w^2_2-(2-\eta)\left(r^4_1+\frac{3}{2}(N^2-1)r^4_2\right)w_3\right
\rbrace\nonumber\ea
We observe the appearance of the coupling $w_3$. This is where the
truncation
of this section becomes relevant: We simply put $w_3=0$ here. We then
observe
that all remaining terms in the curly bracket in (\ref{4.10}) are
positive.
Starting from $w_2(\Lambda)=0$ we conclude that $w_2(k)$ is negative for
all
$k$ between $k_{np}$ and $\Lambda$. In the perturbative region, where
$Ng^2/16\pi^2$ is small, it is easy to see that $w_2$ is of the order
$g^2$. In fact, we may neglect the terms $\sim w^2_2$ and $\sim \eta$ in
the
curly bracket in (\ref{4.10}). One finds for the ratio $\frac{w_2}{g^2}$
an
infrared stable fixpoint
\be\label{4.11}
\frac{\partial}{\partial
t}\left(\frac{w_2}{g^2}\right)=4\frac{w_2}{g^2}+\frac{
127}{360\pi^2}\tau_2 r^{4,2}_0\ee
\be\label{4.12}
w_{2*}(k)=-\frac{127}{1440\pi^2}\tau_2r^{4,2}_0g^2(k)
=-\frac{127}{270}\frac{g^2}{16\pi^2}\ee
which is approached very rapidly. (The last equation in (\ref{4.12}) is
for the choice (\ref{1.2}) and $N=2$.)
It is interesting to insert the value (\ref {4.12}) into the
$\beta$-function for $g^2$ (\ref{4.8}). Expanding in
powers of $g^2$ one has
\ba\label{4.13}
&&\frac{\partial g^2}{\partial
t}=-\frac{22N}{3}\frac{g^4}{16\pi^2}
-\frac{77N^2}{3}\frac{g^6}{(16\pi^2)^2}
+(3N^2-2)\zeta(3)\frac{g^4}{4\pi^2}w_2\nonumber\\
&&=-\frac{22}{3}\frac{g^4N}{16\pi^2}-
\left(\frac{77}{3}+\frac{127}{45}\zeta
(3)\tau_2\left(1-\frac{2}{3N^2}\right)\right)\frac{g^6N^2}{(16\pi^2)^2}
\ea
We note that without the term $\sim w_2$ the coefficient $\sim g^6$
exceeds the perturbative two-loop coefficient
$-\frac{204}{9}\frac{N^2}{(16\pi^2)^2}$
by a little more than 10\%. It is
recomforting to find for the choice (\ref{1.2})
$\left(r_0^{4,2}=\frac{1}{6}
\right)$ the contribution from $w_2$ in
the same order of magnitude as this difference. This gives the hope
that the dominant nonperturbative physics is already contained in
the $F^2$ approximation, whereas additional invariants (like $F^4$ in
this section) give only moderate modifications for $k>k_{np}$. We
emphasize that a full computation of $\beta_{g^2}$ in order $g^6$ should
take additional invariants into account, as for example $(F\tilde F)^2$
or $(D_\mu F^{\mu\nu})^2$. We also observe that for $k=k_{np}$ one has
approximately $Ng^2/16\pi^2=\frac{2}{21}$ such that the validity
of perturbation theory extends roughly to all $k>k_{np}$.

The infrared fixpoint in $w_2/g^2$ implies that the coefficient $W_2$ in
(\ref{4.1}) diverges $\sim k^{-4}$
\be\label{4.14}
W_2\sim -\frac{\bar g^4}{k^4}\ee
Because of the infrared divergence for $k\to 0$, a result of this type
could never have been found in standard perturbation theory. It could,
however, be derived using the
``$\frac{D}{Dt}\approx\frac{\partial}{\partial t}$''-approximation of the
effective average action which we displayed
in eq. (\ref{1.4}). For this purpose one can neglect the $W_k''$ terms on
the r.h.s. of (\ref{1.4}) and approximate $W_k'=w_1$. Then, with
(\ref{2.6})
inserted into (\ref{1.4}), one obtains for the $k$-dependent terms in
$\Gamma_k$
\ba\label{4.15}
\Gamma_k[A,A]&=&\frac{1}{2}{\Tr}_{xcL}\ln\left[w_1{\D}_T+(w_1-
\frac{1}{\alpha_k})D\otimes
D+R_k(w_1{\D}_T+(w_1-\frac{1}{\alpha_k})D\otimes D)\right]\nonumber\\
&&-{\Tr}_{xc}\ln\left[-D^2+R_k(-D^2)\right]\ea
If one extracts the $F^4_{\mu\nu}$-term from these traces, one finds
a (renormalized) coefficient which equals exactly $W_2$,
as extracted from the fixpoint (\ref{4.12}). Clearly
(\ref{4.10}) goes beyond the
$``\frac{D}{Dt}\approx\frac{\partial}{\partial t}''$ approximation. The
terms proportional to $\eta$ and to $w^2_2$ could
not have been obtained in this approximation.

Let us finally consider $k<k_{np}$. The evolution equation for $w_2$
\ba\label{4.16}
&&\frac{\partial}{\partial
t}w_2=4w_2+\frac{g^2(k_{np})}{4\pi^2}\left\lbrace
\tau_2r^{4,2}_0\left(\frac{29}{20}w^2_1-\frac{7}{180}\right)
\right.\nonumber\\
&&+\left.\left(5r^4_1+\frac{9}{2}(N^2-1)r^4_2\right)\frac{w^2_2}{w^4_1}-
\left(r_1^4+\frac{3}{2}(N^2-1)r^4_2\right)
\frac{w_3}{w_1^3}\right\rbrace\ea
again contains only positive terms in the curly bracket if $w_3$ is
neglected and $w^2_1$ is larger than $\frac{7}{261}$. (For
$w_1^2<\frac{7}{261}$
the positive term $\sim w^2_2/w_1^4$ dominates largely.)
The coupling $w_2$, which is negative at the scale $k_{np}$,
will therefore remain negative for $k<k_{np}$.

The investigation of this
section strongly indicates a negative value of the coupling $w_2$ for all
scales. We conclude that the vanishing of $W_1$ at a nonzero confinement
scale $k_\infty$ is unavoidable within our truncation. This is a clear
indication that the $F^2$ term in the effective average action changes
its sign at the scale $k_\infty$!

\section{Gluon condensation}
\setcounter{equation}{0}

If the coefficient $W_1$  in front of the $F_{\mu\nu}F^{\mu\nu}$ term
changes sign for small enough $k$, the solution $A_\mu=0$ cannot
correspond anymore to the minimum of  $\Gamma_k$. Since $\Gamma_k$ is
bounded from below there must be a new solution with
$F_{\mu\nu}F^{\mu\nu}>0$ corresponding  to the  true minimum of $\Gamma_k$
\cite{reuwe}. One therefore  expects a phenomenon of ``gluon
condensation'', i.e. the minimum of $\Gamma_0$ should occur for
nonvanishing $F_{\mu\nu}F^{\mu\nu}$.  In our truncation this suggests that
the minimum of $W_k(\theta)$ should shift to a nonvanishing $\theta_0$ for
small enough $k$. On the other hand, a negative value of $W_2$ excludes a
scenario where the minimum of $W_k$ remains at $\theta=0$ for all
$k>k_\infty$, and then $\theta_0$ continuously moves away from zero for
$k<k_\infty$. A second local minimum of $\Gamma_k$  must already appear
for $k_i>k_\infty$ and become
the absolute minimum for $k<k_c$, with $k_\infty< k_c<k$. At the scale
$k_\infty$ the local minimum at $\theta=0$ finally disappears. This
scenario is closely analogous to a first order phase transition  as a
function of temperature. (In fact, the jump of the minimum of $\Gamma_k$
as a function of $k$  corresponds most probably to a jump of $\Gamma_0$ as
a function of temperature and therefore to a first order phase transition
of the high temperature Yang Mills theory.)
In order to describe such a ``first order scenario'' one has to keep at
least the term $\sim W_3\theta^3$ in a polynomial expansion of $W_k$. We
will show  in this section that a polynomial approximation
\be\label{5.1}
w_k(\vartheta)=w_1\vartheta+\frac{1}{2} w_2\vartheta^2+\frac{1}{6}
w_3\vartheta^3\ee
is indeed sufficient for a qualitative description of gluon condensation.

The first necessary ingredient is a positive coefficient $w_3(k)$. With
(\ref{1.2}) we find in four dimensions the following $\beta$-function
for $k>k_{np}$ (for details see appendix F)
\ba\label{5.2}
&&\frac{\partial}{\partial
t}w_3=(8+\eta)w_3+\frac{g^2}{16\pi^2}\left\lbrace
-\frac{1}{30}\left(\frac{442}{315}-\frac{137}
{210}\eta\right)\tau_3\right.\\
&&\left.+\frac{87}{30}(2-\eta)\tau_2w_2-6\zeta(3)(2-\eta)[(12N^2+23)w^3_2-
(9N^2+8)w_2w_3+N^2w_4]\right\rbrace\nonumber\ea
We neglect the term $\sim w_4$ consistent with our approximation.
For small values of $g^2$ there is an infrared stable fixpoint of
the ratio $w_3/g^2$ corresponding to positive $w_3$
\be\label{5.3}
w_{3*}(k)=\frac{221\tau_3}{37800}\frac{g^2(k)}{16\pi^2}\ee
Actually, all ratios $w_n/g^2$ reach perturbative fixpoints
for $n\geq3$ (cf. appendix F). This justifies the approximation
(\ref{5.1}) at least for small enough $g^2(k)$. (We observe $W_n\sim
g^{2(n-1)}k^{-4(n-1)}$.) A negative value of $w_3(k)$
would require that the positive term $\sim w^3_2$ dominates the
$w_3$-independent part of (\ref{5.2}). This needs $w^2_2>\frac{1}{12N^2
+19}\frac{29\tau_2}{60\zeta(3)}$, or, for $N=2$, $w_2\grgl\frac{1}{8}$.
Such a large value of $w_2$ does not occur for $k>k_{np}$ and we conclude
that
$w_3$ is positive for this range of scales.

For positive $w_3$ the polynomial (\ref{5.1}) is bounded from below for
$\vartheta\geq0$. For $w_1>0$ there is always a local minimum at
$\vartheta
=0$. Two additional extrema are present if
\be\label{5.4} w^2_2>2w_1w_3\ee
There is a minimum for positive $\vartheta_0$
\be\label{5.5}
\vartheta_0=\frac{-w_2+\sqrt{w^2_2-2w_1w_3}}{w_3}\ee
and, for $w_1>0$, a maximum at
\be\label{5.6}
\vartheta_{max}=\frac{-w_2-\sqrt{w_2^2-2w_1w_3}}{w_3}\ee
For $w_1<0$ the origin $\vartheta=0$ turns to a local
maximum. The critical set of couplings where the two minima
are of equal height $(w_k(\vartheta_0)=0)$ corresponds to
\be\label{5.7}
w_1=\frac{3}{8}\frac{w^2_2}{w_3}\ee
For $w_3$ smaller than the critical value, the absolute minimum
occurs at $\vartheta_0$. This corresponds to the phenomenon
of gluon condensation, with Euclidean action in the ground
state given by $w_k(\vartheta_0)$.

For $k>k_{np}$ where $w_1=1$ a quick inspection of the flow equations
(cf. (\ref{4.12}), (\ref{5.3})) shows that $2w_3$ remains larger
than $w^2_2$. There is therefore only one minimum at $\vartheta=0$.
For $k<k_{np}$, however, $w_1$ decreases towards zero. In the region where
$w_3$ remains positive there is necessarily a scale $k_i$ where
(\ref{5.4}) is met and subsequently a scale $k_c$ where $w_1(k_c)$ obeys
(\ref{5.7}). For $k<k_c$ gluon condensation sets in. Since $k_c>k_\infty$
the
establishment of the phenomenon of gluon condensation does not need the
validity of the flow equations down to the scale $k_\infty$. It is
sufficient if the flow equations are a reasonable approximation for
$k\geq k_c$.

What remains to be discussed in our truncation is the positivity
of $w_3$. The relevant flow equation for $k<k_{np}$ reads
\ba\label{5.8}
&&\frac{\partial}{\partial
t}w_3=8w_3+\frac{g^2(k_{np})}{16\pi^2w^2_1}\left
\lbrace-\frac{\tau_3}{30}\left(\frac{137}{105}w^6_1+\frac{31}{315}
w^2_1\right)+\frac{87}{15}\tau_2w^3_1w_2\right.\nonumber\\
&&\left.-12\zeta(3)\left(N^2\frac{w_4}{w_1}-(9N^2+8)\frac{w_2w_3}{w_1^2}+
(12N^2+23)\frac{w_2^3}{w_1^3}\right)\right\rbrace\ea
Neglecting again $w_4$ and for $w_2<0$ the only positive term in the
curly bracket in (\ref{5.8}) is the one $\sim w^3_2/w_1^3$. For
$k\to k_\infty$ this term will finally dominate the flow equation
(\ref{5.8}), drive $w_3$ towards zero, and invalidate the truncation
(\ref{5.1}). We emphasize, however, that the critical condition
(\ref{5.7})
is always met at a scale $k_c$ where $w_3(k_c)$ is positive. Also, from
\ba\label{5.9}
&&\frac{\partial}{\partial
t}w_2=4w_2+\frac{g^2(k_{np})}{16\pi^2w_1^2}\left
\lbrace\tau_2\left(\frac{29}{30}w^2_1-\frac{7}{270}w^4_1\right)\right.
\nonumber\\
&&+\left.\frac{2\zeta(3)}{3}[(9N^2+1)\frac{w^2_2}{w^2_1}-(3N^2-1)
\frac{w_3}{w_1}]\right\rbrace\ea
we learn that the term $\sim w_3$, which was neglected in the previous
section, is actually substantially smaller than the term $\sim w^2_2$
for $k$ between $k_i$ and $k_c$. We therefore see a clear indication
for gluon condensation within the $F^6$ truncation employed in this
section!

The system of the three flow equations (\ref{4.9}), (\ref{5.9}), and
(\ref{5.8}) (with $w_4=0$) could be solved numerically with initial
conditions
given at the scale $k_{np}$. It would be interesting to see if the
expectation value of $\theta$, i.e.
\be\label{5.10}
\theta_0(k)=Z^{-1}_{k_{np}}k^4g^{-2}(k_{np})\vartheta_0(k)\ee
has a tendency to stabilize at a fixed value before $w_1(k)$ or $w_3(k)$
reach zero. The corresponding value could then be taken as a rough
estimate for the size of the gluon condensate. The breakdown of the
approximations of the present paper for $k\to k_\infty$ is, however,
unavoidable for very general reasons that will be discussed in the
final section. This excludes in the present approach a quantitative
determination of the gluon condensate.

\section{Conclusions}
\setcounter{equation}{0}
In this paper we have approximated the exact nonperturbative evolution
equation for Yang-Mills theories \cite{reu} by a truncation where
$\Gamma_k$
is only given as a function $W_k$ of
$\theta=\frac{1}{4}F_{\mu\nu}F^{\mu\nu}$. This ansatz is general enough
to allow for a description of gluon condensation.
A nonvanishing gluon condensate corresponds to an absolute
minimum of $W_k(\theta)$ for $\theta_0>0$ in the limit where the
infrared cutoff $k$ vanishes. We have not solved explicitly for
$k\to0$, but we have found strong indications for gluon condensation
appearing already for nonvanishing $k$: First, the term linear in
$\theta$ in a expansion of $W_k(\theta)$ around $\theta=0$ turns
negative at $k=k_\infty$ (sect. 4). This implies that for $k<k_\infty$
the ``trivial'' configuration $F_{\mu\nu}=0$ cannot correspond to
the minimum of the Euclidean action $\Gamma_k$. Second, we have found
(sect. 5) that a new minimum of $W_k(\theta)$ appears for
$k=k_i$, and that this minimum becomes the absolute minimum of $W_k$ at
the scale $k_c$, with $k_i>k_c>k_\infty$. The truncated set of evolution
equations is qualitatively reasonable for all $k$ down to $k_c$ and we
interpret our results as a strong indication for gluon condensation.

On the other hand, we note that for $k$ in the vicinity of the
confinement scale $k_\infty$ our flow equations become very problematic.
These problems are not originating in the general method but they occur
rather as a particularity of our truncation. There are three different
sorts of problems, and it is instructive to discuss them in detail:

i) The flow equations are not valid for negative $W_k'(\theta)=\partial
W_k/\partial\theta$. For $u=W_k'(\theta)x$ the momentum integrals implied
by the traces of sect. 3, i.e. $\int dxxH(W_k'x)$, are ultraviolet finite
only for $W_k'>0$. In consequence, various terms in the flow equations
(\ref{4.9}), (\ref{5.9}), (\ref{5.8}) diverge for $w_1\to0$. The reason
for this disease is the negative inverse propagator $\Gamma_k^{(2)}\sim
W_k'q^2$ for $W_k'<0$. This implies in particular that the high
momentum modes with $q^2\gg k^2$ are all unstable. Even our optimized
infrared cutoff (\ref{1.2}) cannot remove this problem: The function
$H(u)$ defined in (\ref{3.3}) grows $\sim-u/k^2$ for large $(-u/k^2)$.

The problem described here should, however, be viewed as a pure
artefact of an insufficient truncation. The physics of the high
momentum modes in four-dimensional Yang-Mills theory  should be
governed by asymptotic freedom, and $\Gamma_k$ should be described
by a term $\sim \frac{1}{4} Z_kF_{\mu\nu}F^{\mu\nu}$ with
positive $Z_k$ for all $k$ as far as the high momentum modes are
concerned. In our truncation we have neglected the expected momentum
dependence of $Z_k$. A more reasonable approximation for the term
quadratic in the gauge field $A_\mu$ would be of the sort
\be\label{6.1}
\frac{1}{4}\int d^4xF^z_{\mu\nu}Z_k(-D^2[A])F^{\mu\nu}_z\ee
with $Z_k$ depending on the covariant Laplacian in the adjoint
representation. For the functional form of $Z_k(q^2)$ we expect
for large $q^2$ a $k$-independent positive function $Z(q^2)$. (In
lowest order the logarithmic dependence of
$Z$ on $q^2$ should be determined
by the one-loop $\beta$-function for the running gauge coupling. Indeed,
external momenta of the gluons act as an independent infrared cutoff. For
$k^2\ll q^2$ the running of $Z_k(q^2)$ with $k$ should essentially
stop whereas it is given by the one-loop $\beta$-function
for $k^2\grgl q^2$.)
In contrast, our truncation identifies $Z_k(q^2)$ with a constant
$Z_k(q^2=0)\equiv W_k'(\theta=0)$. This leads to unphysical problems for
negative $W_k'$. In reality, the quantity $Z_k(0)$ may well turn
negative without affecting the high momentum behaviour of $Z_k(q^2)$.
We conclude that with a better truncation the high momentum modes
are always stable if the momentum dependence of the propagators is
properly taken into account. This implies positive $u$ for $x\to\infty$
and no ultraviolet problem can appear for the exponential cutoff
(\ref{1.2}).
The present truncation can nevertheless be employed for all scales $k$
where the inverse propagator for $q^2\approx k^2$ can be approximated by
$Z_k(0)q^2$. One may hope that $Z_k(q^2=k^2)\approx W_k'(\theta=0)$
remains  valid  in the vicinity of $k_c$.

ii) For $k<k_c$ the absolute minimum of $W_k$ occurs for $\theta_0>0$.
For this range of scales it would seem reasonable to expand $W_k'$ around
$\theta_0$ rather than around 0. A polynomial expansion around the true
minimum has turned out to be an important improvement for all models
investigated so far with the help to the average action. The reason
is that the relevant mass terms in the ``ground state'' appear then
directly in the flow equations. Instead, an expansion around
$\theta=0$ involves the ``unphysical mass terms'' corresponding to a
metastable configuration. At the minimum of $W_k$ one always has
\be\label{6.2}
W_k'(\theta_0(k))=0\ee
\be\label{6.3}
W_k''(\theta_0(k))=\bar\lambda>0\ee
Differentiating (\ref{6.2}) with respect to $k$ yields the
running of $\theta_0(k)$, i.e.
\be\label{6.4}
\frac{d\theta_0}{dt}=-\bar\lambda^{-1}\frac{\partial}{\partial t}
W_k'(\theta_0)\ee
This equation, together with flow equations for the couplings
$W^{(n)}(\theta_0)$, could replace the flow equations of the last
section for $k<k_c$.

Unfortunately, a quick inspection of eq. (\ref{3.7}) reveals that the
r.h.s. of such flow equations would diverge in the present truncation.
It is instructive to investigate more in detail the contributions
from the different modes to the evolution equation (\ref{3.4}): First
we have modes with a positive ``transverse kinetic term''
$(W_k'(\theta_0)+B^2W_k''(\theta
_0))x=\bar\lambda B^2x$ that we may call of type I. (Here $x$ denotes
the square of the transverse momentum.)
The contributions
from type I modes as well as from ghosts are unproblematic for $\vartheta
=\vartheta_0$. The type II
modes have an inverse propagator $W_k'(\theta)x+c$ with $c\geq0$.
Contributions from these modes suffer from the ``high momentum disease''
since $W_k'(\theta_0)=0$. The corresponding ultraviolet divergences
in the flow equations could be cured by employing $Z_k(q^2)$ as
discussed above. The infrared behaviour remains without problems.

Finally, there could be
modes of type III with inverse propagator $W_k'(\theta)
x-c,\ c>0$. This would lead to an infrared instability for the modes
with $x\to0$ due to a negative masse like term $-c$. Such an
infrared problem cannot be removed by a modified truncation for the high
momentum modes. It would be an indication that even
at $\theta_0$ the particular configuration with constant
magnetic field introduced in sect. 2 does not correspond to the minimum
of $\Gamma_k$. This is obviously not possible within our truncation
and we find correspondingly that no mode of type III exists in the
spectrum for $\theta=\theta_0$. An expansion around $\vartheta_0$ becomes
therefore
possible for an enlarged truncation of the type (\ref{6.1}).

We should mention that $\Gamma_k$ is, of course, not only defined
for the ground state configuration. Negative eigenvalues of $\Gamma_k
^{(2)}$ for a configuration like (\ref{2.7}) are harmless as long as
$k^2$ is much larger than the absolute value of a negative eigenvalue.
Only once $k^2$ becomes of equal size strong renormalization effects set
in.
They lead to a ``flattening'' of $\Gamma_k$. This is related to the
general property that $\Gamma_k$ becomes a convex quantity for $k\to0$.
We are not dealing with this ``approach to convexity'' \cite{rwy}
in the present investigation.

iii) We observe that the spectrum of small fluctuations around the
constant magnetic field configuration (\ref{2.7}) lacks Euclidean
$SO(4)$ rotation symmetry. This is not surprising since $F_{\mu\nu}$
singles out two of the space directions. In addition, the spectrum is
partially discrete - continuity exists only with respect to the
transversal momentum. The lack of full rotation symmetry and the partial
discreteness of the spectrum are actually related: For an $SO(4)$
symmetric spectrum the continuity in two momentum directions must
extend to all momentum directions. One expects then a spectrum
with a few separate particles, each of them having $\Gamma_k^{(2)}$
depending continuously on a generalized $SO(4)$ invariant of the type
$q_\mu q^\mu$.  The true ground state should
be invariant under a generalized version of Poincar\'e transformations
\cite{reuwe} and therefore have an $SO(4)$ symmetric spectrum. In ref.
\cite{reuwe}
we have proposed a candidate ground state for $d=4$
and gauge group $SU(N),\ N\geq 4$, but
no realistic candidate for the gauge group $SU(3)$ has been found
up to now. It would be very interesting to perform an analysis similar
to the one presented here for this groundstate candidate instead of the
configuration (\ref{2.7}).

In summary, our first attempt to investigate the phenomenon of
gluon condensation with the help of nonperturbative flow equations
is encouraging. Our simple configuration (\ref{2.7}) and the simple
truncation give qualitatively the expected behaviour of $\Gamma_k$:
The minimum of $\Gamma_k$ does not occur for $F_{\mu\nu}=0$ for
sufficiently small $k$, thus indicating the phenomenon of gluon
condensation. The present truncation is, however, insufficient
to describe the true ground state. In consequence, a detailed
quantitative analysis of the size of the gluon condensate seems
premature and should wait for an investigation involving
a configuration which is more similar to the true
ground state.

\section*{Appendix A}

\renewcommand{\theequation}{A.\arabic{equation}}
\setcounter{equation}{0}

In this appendix we collect a few trace identities  which are needed in
order to derive the evolution equation for $W_k$. We choose the
covariantly constant background (\ref{2.7}), (\ref{2.8}). Then eq.
(\ref{2.13}) implies for any function $f$
\be\label{A.1}
\Tr_{xcL}[
P_{\parallel}f(D_\mu,P_{\parallel},P_\perp)]=\Tr_{xL}[f(\partial_\mu,1,0)]
\ee
because $\Tr_c[P_{\parallel}]=n^zn_z=1.$ Writing $P_\perp=1-P_{\parallel}$
and exploiting $ {\cal S}\propto P_{\parallel}$ it is also easy to see
that
\be\label{A.2}
\Tr_{xcL}[P_\perp f(D_\mu,{\cal
S})]=\Tr_{xcL}[f(D_\mu,0)]-\Tr_{xL}[f(\partial_\mu,0)].
\ee
Since  ${\sf F}_{\mu\nu}$ is a constant matrix, the operator $s_{\mu\nu}$
of (\ref{2.14}) commutes with $P_L^{(0)}$ and $P^{(0)}_T$ and satisfies
$\partial^\mu s_{\mu\nu}=0$. This fact can be used to show that
\ba\label{A.3}
&&\Tr_{xL}[f(P_L^{(0)},P^{(0)}_T;s)]\nonumber\\
&=&\Tr_{xL}[f(0,1;s)]+\Tr_x[f(1,0;0)]-\Tr_x[f(0,1;0)]\ea
If one subtracts the same expression with $s=0$ one obtains
\ba\label{A.4}
&&\Tr_{xL}[f(P_L^{(0)},P_T^{(0)};s)-f(P_L^{(0)},P_T^{(0)};0)]\nonumber\\
&=&\Tr_{xL}[f(0,1;s)]-d\Tr_x[f(0,1;0)]
\ea
In the last step we used that $\Tr_{xL}=d\Tr_x$ for an operator
$\sim\delta_{\mu\nu}$. In the above identities the function $f$ may also
depend on further operators provided they commute with those displayed
explicitly and do not introduce any additional colour or Lorentz index
structures.

For the evaluation of $U_1$ in eq. (\ref{B.5}) we need another important
relation:
\be\label{A.5}
\Tr_{xcL}[P_L f ({\cal D}_T)]=\Tr_{xc}[f(-D^2)]
\ee
It follows from the fact that the operator $({\cal
D}_T)_{\mu\nu}=-D^2\delta_{\mu\nu}+2i\bar g F_{\mu\nu}$, when restricted
to the space of longitudinal modes $(a_\mu=(P_L)^\nu_\mu a_\nu)$, has the
same spectrum as $-D^2$ acting on Lorentz scalars. The proof makes
essential use of the identity
\be\label{A.6}
{\cal D}_T(D\otimes D)=(D\otimes D){\cal D}_T=-(D\otimes D)(D\otimes D)
\ee
which holds true whenever the gauge field contained in the covariant
derivatives obeys $D^\mu F_{\mu\nu}=0$.

\section*{Appendix B}

\renewcommand{\theequation}{B.\arabic{equation}}
\setcounter{equation}{0}

In this appendix we use the trace identities derived in appendix A in
order to simplify the flow equation (\ref{3.2}). The first trace on the
r.h.s. of eq. (\ref{3.1}) can be simplified by using the identities of
appendix A. Inserting a factor of  $1=P_{\parallel}+P_\perp$ leads to the
decomposition
\be\label{B.1}
\Tr_{xcL}\left[H(\Gamma^{(2)}_k[A,A])\right]=T^{\parallel}_1+T^\perp_1
\ee
with
\ba\label{B.2}
T^{\pl}_1&=&\Tr_{xcL}\left[P_\pl H(W'_k{\cal
D}_T+[W'_k-\frac{1}{\alpha_k}]D\otimes D+W''_k{\cal S})\right]\nonumber\\
&=&\Tr_{xL}\left[H(-W'_k\partial^2+[W'_k-
\frac{1}{\alpha_k}]\partial\otimes
\partial+W''_k s\right]
\ea
where (\ref{A.1}) was used, and
\ba\label{B.3}
T_1^\perp&=&\Tr_{xcL}\left[P_\perp H(W'_k{\cal
D}_T+[W'_k-\frac{1}{\alpha_k}]D\otimes D+ W''_k{\cal S})\right]\nonumber\\
&=& \Tr_{xcL}\left[ H(W'_k{\cal D}_T+[W'_k-\frac{1}{\alpha_k}] D\otimes
D)\right]\nonumber\\
&&-\Tr_{xL}\left[H(-W'_k\partial^2
+[W'_k-\frac{1}{\alpha_k}]\partial\otimes\partial)\right]
\ea
where (\ref{A.2}) was exploited. Let us write
\be\label{B.4}
\Tr_{xcL}\left[H(\Gamma^{(2)}_k[A,A])\right] =U_1+U_2
\ee
with $U_1$ the ``nonabelian'' trace
\be\label{B.5}
U_1=\Tr_{xcL}\left[ H(W'_k{\cal D}_T+[W'_k-\frac{1}{\alpha_k}]D\otimes
D)\right]\ee
and $U_2$ the sum of $T_1^{\pl}$ and the second term of (\ref{B.3}):
\ba\label{B.6}
U_2&=&\Tr_{xL}\left[H\left(-\partial^2[W_k'
P^{(0)}_T+\frac{1}{\alpha_k}P_L^{(0)}]+W_k'' s\right)\right]\nonumber\\
&&-\Tr_{xL}\left[H\left(-\partial^2[W_k' P^{(0)}_T +\frac{1}{\alpha_k}
P_L^{(0)}]\right)\right]\ea
It is quite remarkable that if we now apply the identity (\ref{A.4}) to
$U_2$, the longitudinal contribution drops out completely and the result
becomes independent of the gauge fixing parameter $\alpha_k$:
\ba\label{B.7}
U_2=\Tr_{xL}[H(-\partial^2 W_k'+W_k'' s)]
-d\Tr_x[H(-\partial^2 W_k')]\ea
The operators entering (\ref{B.7}) are easily diagonalized in a plane-wave
basis. A standard calculation yields, for $W_k'>0,W_k'+B^2W_k''>0$,
\be\label{B.8}
\Omega^{-1}U_2=2v_d\left(\frac{1}{W_k'+B^2
W_k''}-\frac{1}{W_k'}\right)\left(\frac{1}
{W_k'}\right)^{\frac{d}{2}-1}\int^\infty_0 dx \ x^{\frac{d}{2}-1}H(x)\ee
with
$v_d=[2^{d+1}\pi^{d/2}\Gamma(d/2)]^{-1}$. (As always, the argument of
$W_k$ and its derivatives is understood to be $\frac{1}{2} B^2$.)

Next  let us simplify the trace $U_1$ by inserting a pair of projectors:
\ba\label{B.9}
U_1&=&\Tr_{xcL}\left[ P_T H({\cal D}_T[W'_kP_T+\frac{1}{\alpha_k}
P_L])\right]
+\Tr_{xcL}\left[ P_L H ({\cal D}_T[W'_k P_T+\frac{1}{\alpha_k}
P_L])\right]\nonumber\\
&=&\Tr_{xcL}[P_T H(W'_k{\cal D}_T)]+\Tr_{xcL}[P_L
H(\frac{1}{\alpha_k}{\cal D}_T)]\nonumber\\
&=&\Tr_{xcL}[H(W'_k{\cal D}_T)]+\triangle U_1(\alpha_k).\ea
The $\alpha$-dependence of $U_1$ is contained in
\ba\label{B.10}
\triangle U_1(\alpha_k)&=&\Tr_{xcL}\left[ P_L\{ H(\frac{1}{\alpha_k}{\cal
D}_T)-H(W'_k{\cal D}_T)\}\right]\nonumber\\
&=&\Tr_{xc}\left[H(-\frac{1}{\alpha_k} D^2)-H(-W'_kD^2)\right].\ea
In the last line of (\ref{B.10}) we made use of the identity (\ref{A.5}).

By a similar combination of the trace identities we can also evaluate the
last term on the r.h.s. of (\ref{3.2})
\ba\label{B.11}
&&\Tr_{xcL}\left\{P_\perp P_L\left(\tilde H\left(\frac{{\cal
D}_T}{\alpha_k}\right)-H\left(\frac{{\cal
D}_T}{\alpha_k}\right)\right)\right\}\nonumber\\
&=&\Tr_{xc}\left\{\tilde
H\left(-\frac{D^2}{\alpha_k}\right)-H\left(-\frac{D^2}
{\alpha_k}\right)\right\}\nonumber\\
&&-\Tr_x\left\{\tilde
H\left(-\frac{\partial^2}{\alpha_k}\right)-H
\left(-\frac{\partial^2}{\alpha_k}\right)\right\}.\ea
Evaluating the second term in a plane wave basis yields
\be\label{B.12}
\frac{1}{2\Omega}\Tr_x\left\{\tilde
H\left(-\frac{\partial^2}{\alpha_k}\right)-H
\left(-\frac{\partial^2}{\alpha_k}\right)\right\}=v_d\int^\infty_0 dx\
x^{\frac{d}{2}-1}\left(\tilde
H\left(\frac{x}{\alpha_k}\right)-H\left(\frac{x}{\alpha_k}
\right)\right)\ee
At this point we have exploited the various trace identities  as much as
possible.
Combining these results yields the flow equation (\ref{3.4}).

\section*{Appendix C}

\renewcommand{\theequation}{C.\arabic{equation}}
\setcounter{equation}{0}

In this appendix we discuss the group theoretical factors
$\sum_\ell\nu_\ell^{2m}$.  The LHS of the evolution equation is
$\partial_t W_k$. The argument of $W_k$ is $\frac{1}{4} F^z_{\mu\nu}
F_z^{\mu\nu}=\frac{1}{2} B^2$ which is  manifestly independent of the unit
vector $n^z$ which specifies the direction of the field in ``color
space''. The r.h.s. of the evolution equation  consists of expansions such
as (\ref{3.9}) which involve the factors $\sum_\ell\nu^{2m}_\ell$. As
$\{\nu_\ell\}$ are the eigenvalues of $n^zT_z$, we can rewrite them as
\be\label{C.1}
\sum_\ell \nu_\ell^{2m}=n^{z_1} n^{z_2}\cdots n^{z_{2m}} \Tr_c[T_{z_1}
T_{z_2}\cdots T_{z_{2m}}]\ee
where the trace is in the adjoint representation. The question is whether
the invariants (\ref{C.1}) are all independent of the direction of $n^z$.
In appendix D we explain in detail that generically this is \underbar{not}
the case. If the orbit space  of the gauge group in the adjoint
representation is nontrivial, different $n$'s  can lead to different sums
$\sum_\ell \nu^{2m}_\ell$. The resolution to this puzzle is as follows.
For $m=1$ we can use the standard orthogonality relation
\be\label{C.2}
\Tr_c[T_y T_z]=N\delta_{yz}\ee
to prove that $\sum_\ell\nu^2_\ell=N n^z n_z=N$ is independent of the
direction of $n$. Likewise, if the symmetric invariant tensor $\Tr_c\left[
T_{(z_1}\cdots T_{z_{2m})}\right]$ is proportional to the trivial one,
$\delta_{(z_1z_2}\delta_{z_3z_4}\cdots\delta_{z_{2m-1}z_{2m})}$, we can
again use the normalization condition $n^zn_z=1$ to show that the r.h.s.
of (\ref{C.1}) is independent of $n$. The situation changes if there
exists a totally symmetric invariant tensor ${\cal T}_{z_1z_2\cdots
z_{2m}}$ which is different from the trivial one. Then we might have
\be\label{C.3}
\Tr_c\left[ T_{(z_1}\cdots T_{z_{2m})}\right]=\tau_m\delta_{(z_1z_2}\cdots
\delta_{z_{2m-1}z_{2m})}+{\cal T}_{z_1z_2\cdots z_{2m}}\ee
with some coefficient $\tau_m$. (If there exists more than one ${\cal T}$
an appropriate sum is implied.) In general $n^{z_1} n^{z_2}...{\cal
T}_{z_1z_2\cdots}$ will be direction dependent \cite{rai}.
If some invariant tensor ${\cal T}$ exists, the correct way of deriving
the evolution equation for $W_k$ is to compare coefficients of a fixed
tensor structure on both sides of the equation. Clearly the l.h.s.,
$\partial
_tW_k(\frac{1}{4}F^2_{\mu\nu})$, gives rise to the trivial tensor
structure
only. Therefore only the $\tau_m$-piece of (\ref{C.3}) should be
kept in (\ref{C.1}) and the ($n$-dependent) part coming from ${\cal T}$
has to be discarded. Thus (\ref{3.9}) may be used on the r.h.s. of the
equation for $W_k$ provided we interpret $\sum \nu^{2m}_l$ as the
coefficient $\tau_m$. On the other hand, nontrivial ${\cal T}$ permits
to construct additional invariants from an even number of covariantly
conserved $F_{\mu\nu}$. The truncation $W(\theta)$ is not sufficient
anymore to parametrize the most general effective action for constant
magnetic fields of the type introduced in sect. 2 (with covariantly
constant $F_{\mu\nu}$). The evolution equation for the new invariants
can now be extracted by projecting the r.h.s. on the appropriate tensor
structure. We will not pursue this generalization in the present paper.

In appendix D we show that for $SU(2)$ this complication is
absent. There exists no additional invariant tensor ${\cal T}$, and one
finds the $n$-independent result
\be\label{C.4}
\sum^3_{l=1} \nu_l^{2m}=2\ee
for all $m=1,2...$.

\section*{Appendix D}

\renewcommand{\theequation}{D.\arabic{equation}}
\setcounter{equation}{0}

In this appendix we investigate in more detail the group-theoretical
quantities $\sum_l\nu_l^{2m}$ which occur in many calculations involving
covariantly
constant backgrounds of the type $A_\mu^z=n^z{\sf A}_\mu$. Here we
consider
an arbitrary (semi-simple, compact) gauge group $G$ with structure
constants
$f^{wyz}$. For a fixed unit vector $n^z$ we consider the matrix
\be\label{D.1}
\hat n^{yz}=n^w(T^w)^{yz}=if^{ywz}n^w\ee
The numbers $\nu_l,l=1,...,dim\ G$ are the eigenvalues of $\hat n$:
$\hat n^{yz}\psi^z_l=\nu_l\psi^y_l$. This equation can be rewritten in a
more
suggestive form. Let $t^z$ denote the generators of $G$ in an arbitrary
representation: $[t^w,t^y]=if^{wyz}t^z$. If we define
\be\label{D.2}
\tilde n=n^zt^z,\quad\tilde\psi_l=\psi^z_lt^z\ee
the eigenvalue equation becomes
\be\label{D.3}
[\tilde n,\tilde\psi_l]=\nu_l\tilde\psi_l\ee
Clearly the $\nu_l$'s do not depend on the representation chosen. We would
like to know how the spectrum $\lbrace\nu_l\rbrace$ depends on the
vector $n$. First of all, it is clear that if $V$ is any group element
in the $t$-representation, the matrices $\tilde n$ and $\tilde n'=
V\tilde nV^{-1}$ have
the same spectrum, i.e. the spectrum is constant along the orbit of $G$ in
the adjoint representation. If two directions $n$ and $n'$ are not
related by a group transformation, then the spectra
$\lbrace\nu_l(n)\rbrace$
and $\lbrace\nu_l(n')\rbrace$ can be different. Typically, for $G$ large
enough \cite{rai}, the orbit space is indeed nontrivial, and the spectrum
``feels'' the direction of $n^z$.

Let us go over from the basis $\lbrace T^z\rbrace$ to the Cartan-Weyl
basis $\lbrace H_i,E_{\vec\alpha}\rbrace$ of the abstract Lie algebra.
Here ${\vec\alpha}\in {\sf I\!R}^r$ are the root vectors and
$i=1,...,r\equiv\ rank\ G$. For definiteness we assume that the
$t^z$'s are in the fundamental representation where we write $\lbrace
h_i,e_{\vec\alpha}\rbrace$ for the Cartan-Weyl basis. Thus
\be\label{D.4}
[H_i,E_{\vec \alpha}]=\alpha_iE_{\vec\alpha}\quad{\rm and}\quad
[h_i,e_{\vec\alpha}]=\alpha_ie_{\vec\alpha}\ee
We assume that the generators $h_i$ of the Cartan subalgebra are given by
diagonal matrices. By an appropriate transformation $\tilde n\to V\tilde n
V^{-1}$ any $\tilde n$ can be brought to diagonal form. Therefore, in
order
to investigate  the $n$-dependence of $\lbrace\nu_l(n)\rbrace$, it is
sufficient to consider $\tilde n$'s which are in the Cartan subalgebra:
$\tilde n=\sum_{i=1}^rn_ih_i$. For this choice
\be\label{D.5}
[\tilde
n,e_{\vec\alpha}]=\left(\sum^r_{i=1}n_i\alpha_i\right)e_{\vec\alpha},
\quad[\tilde n,h_i]=0\ee
and the nonvanishing eigenvalues $\nu_l=\nu_{\vec\alpha}$ are given by
\be\label{D.6}
\nu_{\vec\alpha}=\sum^r_{i=1}n_i\alpha_i\ee
Therefore the quantities $\sum_l\nu_l^{2m}$ can be computed explicitly
from the root system:
\be\label{D.7}
\sum^{dim G}_{l=1}\nu^{2m}_l=\sum_{roots\{\vec\alpha\}}\left(
\sum^{rank\ G}_{i=1}n_i\alpha_i\right)^{2m}\ee
Let us consider a few simple examples. For $G=SU(2)$ we have
%r=1,n_1-1$ and there are only three (one-component) roots:
$\alpha=\pm1$. Thus
\be\label{D.8}
\sum^3_{l=1}\nu_l^{2m}=2,\quad m=1,2,3,... \ee
depends neither on the direction $n^z$ nor on the power $m$. This
degeneracy can be understood by noting that for $SU(2)$ the square of
the matrix (\ref{D.1}) is the projector $P_\perp$: $\hat n\hat n
=P_\perp$. This means that $\sum_l\nu^{2m}_l=\Tr(P^m_{\perp})=\Tr
(P_\perp)=2$, as it should be. Contracting eq. (\ref{C.3}) with
$n^{z_1}...n^{z_{2m}}$ and comparing the result to (\ref{D.8}) we see
that there exists no nontrivial invariant tensor ${\cal T}$ and
that $\tau_m=2$ for all $m$.

For $G=SU(3)$ we have $r=2$, and a 2-component unit vector $(n_1,n_2)$
specifies the direction of the field in the Cartan subalgebra. Using
the explicit form of the roots it is straightforward to derive that
\be\label{D.9}
\sum^8_{l=1}\nu^{2m}_l=2^{1-2m}\left[(n_1+\sqrt
3n_2)^{2m}+(n_1-\sqrt3n_2)^{2m}+(2n_1)^{2m}\right]\ee
For $m=1$ and $m=2$ it turns out that this expression depends on $n_1$
and $n_2$ only via $n^2_1+n^2_2=1$, and one obtains the
direction-independent
results
\be\label{D.10}
\sum^8_{l=1}\nu^2_l=3,\quad\sum^8_{l=1}\nu^4_l=\frac{9}{4}\ee
Starting from $m=3$, the invariants are explicitly $n$-dependent. Writing
$n_1=\cos\theta, n_2=\sin\theta$ we find for $m=3$\ba\label{D.11}
\sum_{l=1}^8\nu_l^6=\frac{3}{16}\left[11\cos^6\theta+15\cos^4\theta\sin^2
\theta
+45\cos^2\theta\sin^4\theta+9\sin^6\theta\right]\ea
As discussed in section 3, the $n$-dependence is related to the
existence of a nontrivial invariant tensor ${\cal T}_{z_1...z_6}$.
However, we are not going to calculate the corresponding coefficient
$\tau_3$ here.

\section*{Appendix E}

\renewcommand{\theequation}{E.\arabic{equation}}
\setcounter{equation}{0}

In the regime where $\bar gB/k^2\grgl 1$ the use of the Euler-McLaurin
 expansion (\ref{3.9}) becomes questionable and we should look for an
alternative representation of the spectral sums (\ref{3.7}) and
(\ref{3.8}).
In this section we use the Schwinger proper-time representation
\cite{direu}.
It can be easily applied only for the simplified cutoff function
\be\label{E.1} R_k(x)=Z_kk^2.\ee
In this case one may write $(x\equiv {\cal D}_T)$
\be\label{E.2}
H(x)\equiv\frac{\partial_tR_k(x)}{x+R_k(x)}=\frac{\partial}{\partial
t}(Z_kk^2)\int^\infty_0dse^{-sZ_kk^2}e^{-sx}\ee
Inserting this representation into (\ref{3.7}) we may employ
\[\sum^\infty_{n=0}\exp[-sW_k'\bar g|\nu_l|B(2n+1)]=\frac{1}
{2\sinh[sW_k'\bar g|\nu_l|B]}\]
and
\[\int^\infty_0dxx^{\frac{d}{2}-2}e^{-sW_k'x}=v^{-1}_{d-2}2^{1-d}
\pi^{1-\frac{d}{2}}(sW_k')^{1-\frac{d}{2}}\]
as long as $W_k'>0$. One finds
\ba\label{E.3}
&&\Omega^{-1}\Tr_{xcL}[H(W_k'\D_T)]=2(4\pi)^{-\frac{d}{2}}(W_k')^{1
-\frac{d}{2}}\nonumber\\
&&\cdot\frac{\partial}{\partial t}(Z_kk^2)\sum^{N^2-1}_{l=1}\bar g|\nu_l|B
\int^\infty_0\frac{ds}{s}s^{(4-d)/2}e^{-sZ_kk^2}\nonumber\\
&&\cdot\left[\frac{d}{2\sinh(sW_k'\bar g|\nu_l|B)}-\exp(-sW_k'\bar
g|\nu_l|B)
+\exp(+sW_k'\bar g|\nu_l|B)\right]\ea
where the last exponential is due to the unstable
mode. For $Z_kk^2<W_k'\bar g
|\nu_l|B$ it makes the $s$-integration divergent at the upper (i.e. IR)
limit. In conventional calculations of the one-loop effective action
this creates a problem from the outset, because one attempts to work at
$k^2\to0$ there. In the present formulation everything is well defined
for $k^2$ sufficiently large, and one interesting question is how the
renormalization group flow behaves as one approaches $k^2\approx \theta$
from above.

In the UV limit $s\to 0$ the terms inside the square bracket
in (\ref{E.3}) behave as
\be\label{E.4}
[...]=\frac{d}{2\bar g|\nu_l|BW_k'(B^2/2)}\cdot\frac{1}{s}+O(s)\ee
While the $O(s)$-terms do not lead to UV-divergences for $d<6$, the
term $\sim 1/s$ leads to a divergent contribution to the proper-time
integral. Though the factor $\bar g|\nu_l|B$ cancels against a similar one
coming from the density of states, this divergent piece is still
field-dependent because of the $B$-dependence of $W_k'$.  This
UV divergence shows a failure of the truncation for
the mass-type cutoff function $R_k=Z_kk^2$. The latter may
be used only together with the approximation $W_k''=0$
on the r.h.s. of the flow equation. The divergent
piece in the proper-time integral is an irrelevant constant then.

Using (\ref{E.3}) and a similar formula for the scalar traces in eq.
(\ref{3.4}), we obtain (up to an irrelevant constant and for $\tilde Z
_k=1/\alpha_k,\tilde\eta=0$)
\ba\label{E.5}
&&\frac{\partial}{\partial t}W_k(\frac{1}{2}B^2)=(4\pi)^{-d/2}\sum^{N^2-1}
_{l=1}\bar g|\nu_l|Bk^2\int^\infty_0ds\ s^{1-d/2}\cdot\nonumber\\
&&\cdot\left\lbrace(2-\eta)Z_k(W_k')^{1-\frac{d}{2}}\left[\frac{d-1}
{2\sinh(sW_k'\bar g|\nu_l|B)}+2\sinh(sW_k'\bar g|\nu_l|B)\right]
e^{-sZ_kk^2}\right.
\nonumber\\
&&\left.-\frac{\exp(-sk^2)}{\sinh(s\bar g
|\nu_l|B)}\right\rbrace\ea
This evolution equation is the analogue of (\ref{3.11}) with the
additional
assumption $W_k''=0$. Contrary to the Euler-McLaurin series it is valid
even
for strong fields $\bar gB\approx k^2$.

For $\bar gB\ll k^2$ the r.h.s. of (\ref{E.5}) can be expanded in powers
of B. Apart from the different form of $R_k$, this reproduces the
Euler-McLaurin
expansion. Expanding up to order $B^4$ we find, for instance,
\be\label{E.6}
\frac{\partial}{\partial t}\left(\frac{w_2}{g^2}\right)=4\frac{w_2}{g^2}
+\frac{127}{360\pi^2}\tau_2r_0^{4,2}\ee
with $r_0^{4,2}=2$.
This result is the counterpart of eq. (\ref{4.11}) which had been obtained
with the exponential cutoff for which $r_0^{4,2}=1/6$. In accordance with
the
(\ref{3.13}) we find an additional factor of 12 in the second term on the
r.h.s. of (\ref{E.6}). Hence also the value of the fixpoint $w_{2*}(k)$
is 12 times larger than the result (\ref{4.12}). In view of the discussion
following (\ref{4.13}) this means that for the mass-type cutoff
$R_k=Z_kk^2$
this higher-order correction is much larger than for the exponentially
decreasing $R_k$ of (\ref{1.2}). This is probably closely related to
the ultraviolet problems and
indicates that, though computationally
more difficult to handle, the cutoff (\ref{1.2}) should be used for
reliable estimates.

\section*{Appendix F}

\renewcommand{\theequation}{F.\arabic{equation}}
\setcounter{equation}{0}

In this appendix we derive the flow equation in the $F^6$ truncation.
We start from the evolution equation for $\ddot w(\vartheta)$ which
follows
from differentiating (\ref{4.7}) with respect to $\vartheta$
\ba\label{F.1}
&&\frac{\partial}{\partial t}\ddot w=(4+\eta)\ddot w+4\vartheta w^{(3)}
\nonumber\\
&&-(2-\eta)v_dg^2\dot w^{-\frac{d}{2}}\left\lbrace(d-2)\sum^\infty_{m=1}
\tau_m(C^d_m-E_m)r^{d,m}_0\right.\nonumber\\
&&\left[(2\vartheta\dot w^2)^{m-1}\left(((8-d)m-d)\dot w\ddot
w+((4-2d)m-d+
\frac{d^2}{2})\vartheta\ddot w^2+(4m-d)\vartheta\dot w
w^{(3)}\right)\right.
\nonumber\\
&&+2(m-1)(2\vartheta\dot w^2)^{m-2}\dot w^2(\dot w+2\vartheta\ddot
w)(2m\dot w+(4m-d)\vartheta\ddot w)\Biggr]\nonumber\\
&&+r^d_1\left[\frac{4w^{(3)}+2\vartheta w^{(4)}}{\dot w+2\vartheta \ddot
w}-
\frac{4(\ddot w+\vartheta w^{(3)})(3\ddot w+2\vartheta
w^{(3)})+2\vartheta\ddot w(5w^{(3)}+2\vartheta w^{(4)})}{(\dot
w+2\vartheta\ddot w)^2}\right.\nonumber\\
&&+\frac{4\vartheta\ddot w(3\ddot w+2\vartheta w^{(3)})^2}{(\dot
w+2\vartheta
\ddot w)^3}\nonumber\\
&&\left.-d\frac{2\ddot w^2+3\vartheta\ddot w w^{(3)}}{\dot w(\dot
w+2\vartheta
\ddot w)}+\frac{3}{2}d\frac{\vartheta\ddot w^3}{\dot w^2(\dot
w+2\vartheta\ddot w)}+2d\frac{\vartheta\ddot w^2(3\ddot w+2\vartheta
w^{(3)})}{\dot w(\dot w+2\vartheta\ddot w)^2}\right]\nonumber\\
&&\left.+\frac{1}{4}d(d-1)(d-2)(N^2-1)r^d_2\left[\frac{w^{(3)}}{\dot w}-
\frac{d+2}{2}\frac{\ddot w^2}{\dot w^2}\right]\right\rbrace\nonumber\\
&&+8(d-2)v_dg^2\sum^\infty_{m=1}m(m-1)
\tau_mE_mr^{d,m}_0(2\vartheta)^{m-2}\ea
Taking one further $\vartheta$-derivative at $\vartheta=0$ we obtain
for $d=4$ and $k>k_{np}$ the flow equation for $w_3$ (with $C^4_2-E_2=
-\frac{29}{80},\ C^4_3-E_3=-\frac{137}{10080}$ and $E_3=\frac{31}{30240})$
\ba\label{F.2}
&&\frac{\partial}{\partial t}w_3=(8+\eta)w_3\nonumber\\
&&+\frac{g^2}{16\pi^2}\left\lbrace\left(\frac{442}{315}-\frac{137}{210}
\eta\right)\tau_3r^{4,3}_0+\frac{87}{5}(2-\eta)\tau_2r^{4,2}_0w_2
\right.\nonumber\\
&&-(2-\eta)r^4_1(3w_4-51w_2w_3+105w^3_2)\nonumber\\
&&-3(2-\eta)(N^2-1)r^4_2(w_4-9w_2w_3+12w^3_2)\Biggr\rbrace\ea
For the infrared cutoff (\ref{1.2}) one has $r^{4,2}_0=\frac{1}{6},
r_0^{4,3}
=-\frac{1}{30}$ and we observe that the perturbatively leading term
$\sim g^2$ is negative. This implies a positive perturbative fixpoint
value
for the ratio $w_3/g^2$.

This discussion can easily be generalized for
arbitrary $w_n$. In lowest order in $g^2$ the flow equation (\ref{F.1})
simplifies considerably $(k>k_{np})$
\ba\label{F.3}
&&\frac{\partial}{\partial t}\ddot w=4\ddot w+4\vartheta
w^{(3)}\nonumber\\
&&-8(d-2)v_dg^2\sum^\infty_{m=2}m(m-1)\tau_mr^{d,m}_0
\left(C^d_m-2E_m\right)
(2\vartheta)^{m-2}\ea
This implies for $n\geq 2$ the flow equations
\be\label{F.4}
\frac{\partial}{\partial t}w_n=4^{n-1}w_n-(d-2)v_dg^22^{n+1}n!\tau_n
r^{d,n}_0\left(C^d_n-2E_n\right)\ee
and the infrared fixpoint values
\be\label{F.5}
\left(\frac{w_n}{(d-2)v_dg^2}\right)_*=2^{3-n}n!\tau_nr^{d,n}_0\left(C^d_n
-2E_n\right)\ee
For $d=4$ and with (\ref{3.14}) this yields
\ba\label{F.6}
w_{n*}&=&d_n\frac{g^2}{16\pi^2}\nonumber\\
d_n&=&\frac{2^{4-n}n!}{(2n-1)!}\tau_nB_{2n-2}\left(\frac{2^{2n-1}-1}
{2n}B_{2n}-1\right)\ea
where for $SU(2)$
\be\label{F.7}
\frac{2d_2}{\tau_2}=-\frac{127}{270},
\quad\frac{2d_3}{\tau_3}=\frac{221}{37800}
\ee
The series of $d_n$ is alternating as long as the bracket is dominated
by -1.

Finally, for $k<k_{np}$ the flow equation for $w_3$ follows from
(\ref{F.1})
as
\ba\label{F.8}
&&\frac{\partial}{\partial t}w_3=8w_3+\frac{g^2}{16\pi^2w^2_1}\left\lbrace
\tau_3r_0^{4,3}\left(\frac{137}{105}w_1^6+\frac{31}{315}w^2_1\right)
+\frac{174}{5}\tau_2r^{4,2}_0w^3_1w_2\right.\nonumber\\
&&-6r_1^4\left(\frac{w_4}{w_1}-17\frac{w_2w_3}{w^2_1}
+35\frac{w^3_2}{w_1^3}\right)\nonumber\\
&&\left.-6(N^2-1)r^4_2\left(\frac{w_4}{w_1}-9\frac{w_2w_3}
{w_1^2}+12\frac{w^3_2}{w_1^3}\right)\right\rbrace\ea


\begin{thebibliography}{12}
\bibitem{sav}
G. K. Savvidy, Phys. Lett. {\bf 71B} (1977) 133
\bibitem{dit} W. Dittrich and M. Reuter, Phys. Lett. {\bf B128} (1983)
321;\\
M. Reuter and W. Dittrich, Phys. Lett. {\bf B144} (1984) 99
\bibitem{Av} C. Wetterich, Nucl. Phys. {\bf B352} (1991) 529;\\
C. Wetterich, Z. Phys. {\bf C57} (1993) 451
\bibitem{Ev} C. Wetterich, Phys. Lett. {\bf B301} (1993) 90
\bibitem{Rg} F. Wegner, A. Houghton, Phys. Rev. {\bf A8} (1973) 401;\\
K. G. Wilson, I. G. Kogut, Phys. Rep. {\bf 12} (1974) 75;\\
S. Weinberg, {\it Critical Phenomena for Field Theories}, Erice
Subnucl. Phys. 1 (1976);\\
J. Polchinski, Nucl. Phys. {\bf B231} (1984) 269;\\
A. Hasenfratz, P. Hasenfratz, Nucl. Phys. {\bf 270} (1986) 685
\bibitem{Mw} M. Bonini, M. D'Attanasio, and G. Marchesini, Parma
preprint URPF 92-360;\\
C. Wetterich, Int. J. Mod. Phys. {\bf A9} (1994) 3571
\bibitem{reu}  M. Reuter and C. Wetterich, Nucl. Phys. {\bf B417} (1994)
181
\bibitem{Gau} M. Bonini, M. D'Attanasio, and G. Marchesini,
Nucl. Phys. {\bf B418} (1994) 81; {\bf B421} (1994) 429;\\
U. Ellwanger, Heidelberg preprint HD-THEP-94-2
\bibitem{reuwe} M. Reuter and C. Wetterich, Phys. Lett. {\bf B334} (1994)
412
\bibitem{abb} L. F. Abbott, Nucl. Phys. {\bf B185} (1981) 189;\\
W. Dittrich and M. Reuter, {\it Selected Topics in Gauge Theories},
Lecture Notes in Physics, Vol. 244, Springer, Berlin, 1986
\bibitem{wet}
M. Reuter and C. Wetterich, Nucl. Phys. {\bf B391} (1993) 147; Nucl.
Phys. {\bf B408} (1993) 91
\bibitem{rwe} M. Reuter and C. Wetterich, Nucl. Phys. {\bf B427} (1994)
291
\bibitem{adl}
S. L. Adler, Phys. Rev. {\bf D23} (1981) 2905;
Nucl. Phys. {\bf B217} (1983) 381;\\
S. L. Adler and T. Piran, Phys. Lett. {\bf 113B} (1982) 405; {\bf 117B}
(1982) 91
\bibitem{nol}
N. K. Nielsen and P. Olesen, Nucl. Phys. {\bf B144} (1978) 376; Phys.
Lett. {\bf 79B} (1978) 304
\bibitem{ksw} H. J. Kaiser, K. Scharnhorst, and E. Wieczorek,
J. Phys. {\bf G16} (1990) 161
\bibitem{direu} W. Dittrich and M. Reuter, {\it Effective
Lagrangians in Quantum Electrodynamics}, Lecture Notes in Physics,
Vol. 220, Springer, Berlin, 1985
\bibitem{zuk}
J. A. Zuk, J. Phys. A: Math. Gen. {\bf 18} (1985) 1795;\\
C. M. Fraser, Z. Phys. {\bf C28} (1985) 101;\\
R. I. Nepomechie, Phys. Rev. {\bf D31} (1985) 3291
\bibitem{mgs}
M. G. Schmidt and C. Schubert, Phys. Lett. {\bf B318} (1993) 438,
Phys. Lett. {\bf B331} (1994) 69;\\
D. Fliegner, M. G. Schmidt, and C. Schubert, Z. Phys. {\bf C64} (1994) 111
\bibitem{rwy} L. O'Raifeartaigh, A. Wipf, and H. Yoneyama, Nucl. Phys.
{\bf B271} (1986) 653;\\
A. Ringwald and C. Wetterich, Nucl. Phys. {\bf B334} (1990) 506;\\
N. Tetradis and C. Wetterich, Nucl. Phys. {\bf B383} (1992) 197
\bibitem{rai}
L. O'Raifeartaigh, {\it Group structure of gauge theories}, Cambridge
University Press, Cambridge, 1986

\end{thebibliography}
\end{document}